\pdfoutput=1
\documentclass[12pt,a4paper]{article}

\usepackage{ifthen} 
\newboolean{pdflatex}
\setboolean{pdflatex}{true} 

\newboolean{articletitles}
\setboolean{articletitles}{true} 

\newboolean{uprightparticles}
\setboolean{uprightparticles}{false} 

\newboolean{inbibliography}
\setboolean{inbibliography}{false} 

\def\paperauthors{LHCb collaboration} 
\def\paperasciititle{Measurement of the CP asymmetry in B- to Ds-Dz and B- to D-Dz decays} 
\def\papertitle{Measurement of the \CP asymmetry in \decay{\Bm}{\Dsm\Dz} and \decay{\Bm}{\Dm\Dz} decays} 
\def\paperkeywords{{High Energy Physics}, {LHCb}} 
\def\papercopyright{CERN on behalf of the LHCb collaboration}
\def\paperlicence{CC-BY-4.0}
\def\paperlicenceurl{https://creativecommons.org/licenses/by/4.0/}

\usepackage[top=1in, bottom=1.25in, left=1in, right=1in]{geometry}

\usepackage{microtype}
\usepackage{lineno}  
\usepackage{xspace} 
\usepackage{caption} 

\usepackage{graphicx}  
\usepackage{color}
\usepackage{colortbl}
\graphicspath{{./figs/}} 

\usepackage{amsmath} 
\usepackage{amssymb}
\usepackage{amsfonts}
\usepackage{upgreek} 

\newcommand*\patchAmsMathEnvironmentForLineno[1]{%
\expandafter\let\csname old#1\expandafter\endcsname\csname #1\endcsname
\expandafter\let\csname oldend#1\expandafter\endcsname\csname
end#1\endcsname
 \renewenvironment{#1}%
   {\linenomath\csname old#1\endcsname}%
   {\csname oldend#1\endcsname\endlinenomath}%
}
\newcommand*\patchBothAmsMathEnvironmentsForLineno[1]{%
  \patchAmsMathEnvironmentForLineno{#1}%
  \patchAmsMathEnvironmentForLineno{#1*}%
}
\AtBeginDocument{%
\patchBothAmsMathEnvironmentsForLineno{equation}%
\patchBothAmsMathEnvironmentsForLineno{align}%
\patchBothAmsMathEnvironmentsForLineno{flalign}%
\patchBothAmsMathEnvironmentsForLineno{alignat}%
\patchBothAmsMathEnvironmentsForLineno{gather}%
\patchBothAmsMathEnvironmentsForLineno{multline}%
\patchBothAmsMathEnvironmentsForLineno{eqnarray}%
}

\usepackage{hyperxmp}

\usepackage[pdftex,
            pdfauthor={\paperauthors},
            pdftitle={\paperasciititle},
            pdfkeywords={\paperkeywords},
            pdfcopyright={Copyright (C) \papercopyright},
            pdflicenseurl={\paperlicenceurl}]{hyperref}

\usepackage[all]{hypcap} 

\usepackage{xspace} 
\usepackage{upgreek}

\def\lhcb {\mbox{LHCb}\xspace}

\def\MagUp {\mbox{\em Mag\kern -0.05em Up}\xspace}

\ifthenelse{\boolean{uprightparticles}}%
{

 \def\Pmu         {\ensuremath{\upmu}\xspace}                 
 \def\Pnu         {\ensuremath{\upnu}\xspace}                 
                  
 \def\Ppi         {\ensuremath{\uppi}\xspace}

 \def\PDelta      {\ensuremath{\Delta}\xspace}                 
 \def\PXi      {\ensuremath{\Xi}\xspace}                 
 \def\PLambda      {\ensuremath{\Lambda}\xspace}                 
 \def\PSigma      {\ensuremath{\Sigma}\xspace}                 
 \def\POmega      {\ensuremath{\Omega}\xspace}                 
 \def\PUpsilon      {\ensuremath{\Upsilon}\xspace}                 
 
 \def\PB      {\ensuremath{\mathrm{B}}\xspace}                 
                  
 \def\PD      {\ensuremath{\mathrm{D}}\xspace}

 \def\PK      {\ensuremath{\mathrm{K}}\xspace}

 \def\Pb      {\ensuremath{\mathrm{b}}\xspace}                 
 \def\Pc      {\ensuremath{\mathrm{c}}\xspace}                 
 \def\Pd      {\ensuremath{\mathrm{d}}\xspace}

 \def\Pi      {\ensuremath{\mathrm{i}}\xspace}

 \def\Ps      {\ensuremath{\mathrm{s}}\xspace}

}
{

 \def\Pmu         {\ensuremath{\mu}\xspace}                 
 \def\Pnu         {\ensuremath{\nu}\xspace}                 
                  
 \def\Ppi         {\ensuremath{\pi}\xspace}

 \mathchardef\PDelta="7101
 \mathchardef\PXi="7104
 \mathchardef\PLambda="7103
 \mathchardef\PSigma="7106
 \mathchardef\POmega="710A
 \mathchardef\PUpsilon="7107
                  
 \def\PB      {\ensuremath{B}\xspace}                 
                  
 \def\PD      {\ensuremath{D}\xspace}

 \def\PK      {\ensuremath{K}\xspace}

 \def\Pb      {\ensuremath{b}\xspace}                 
 \def\Pc      {\ensuremath{c}\xspace}                 
 \def\Pd      {\ensuremath{d}\xspace}

 \def\Pi      {\ensuremath{i}\xspace}

 \def\Ps      {\ensuremath{s}\xspace}

}

\makeatletter
\ifcase \@ptsize \relax
  \newcommand{\miniscule}{\@setfontsize\miniscule{4}{5}}
\or
  \newcommand{\miniscule}{\@setfontsize\miniscule{5}{6}}
\or
  \newcommand{\miniscule}{\@setfontsize\miniscule{5}{6}}
\fi
\makeatother

\DeclareRobustCommand{\optbar}[1]{\shortstack{{\miniscule (\rule[.5ex]{1.25em}{.18mm})}
  \\ [-.7ex] $#1$}}

\def\mun        {{\ensuremath{\Pmu^-}}\xspace} 

\def\neub       {{\ensuremath{\overline{\Pnu}}}\xspace}

\def\neumb      {{\ensuremath{\neub_\mu}}\xspace}

\def\dquark    {{\ensuremath{\Pd}}\xspace}

\def\squark    {{\ensuremath{\Ps}}\xspace}

\def\cquark    {{\ensuremath{\Pc}}\xspace}

\def\bquark    {{\ensuremath{\Pb}}\xspace}

\def\pion   {{\ensuremath{\Ppi}}\xspace}

\def\pip    {{\ensuremath{\pion^+}}\xspace}
\def\pim    {{\ensuremath{\pion^-}}\xspace}

\def\kaon    {{\ensuremath{\PK}}\xspace}
  \def\Kbar    {{\kern 0.2em\overline{\kern -0.2em \PK}{}}\xspace}

\def\KorKbar    {\kern 0.18em\optbar{\kern -0.18em K}{}\xspace}

\def\Kp      {{\ensuremath{\kaon^+}}\xspace}
\def\Km      {{\ensuremath{\kaon^-}}\xspace}

\def\KS      {{\ensuremath{\kaon^0_{\mathrm{ \scriptscriptstyle S}}}}\xspace}

  \def\Dbar    {{\kern 0.2em\overline{\kern -0.2em \PD}{}}\xspace}
\def\D       {{\ensuremath{\PD}}\xspace}

\def\DorDbar    {\kern 0.18em\optbar{\kern -0.18em D}{}\xspace}
\def\Dz      {{\ensuremath{\D^0}}\xspace}
\def\Dzb     {{\ensuremath{\Dbar{}^0}}\xspace}
\def\Dp      {{\ensuremath{\D^+}}\xspace}
\def\Dm      {{\ensuremath{\D^-}}\xspace}

\def\Dstarz  {{\ensuremath{\D^{*0}}}\xspace}

\def\Dstarp  {{\ensuremath{\D^{*+}}}\xspace}

\def\Dsm     {{\ensuremath{\D^-_\squark}}\xspace}

\def\B       {{\ensuremath{\PB}}\xspace}
\def\Bbar    {{\ensuremath{\kern 0.18em\overline{\kern -0.18em \PB}{}}}\xspace}

\def\BorBbar    {\kern 0.18em\optbar{\kern -0.18em B}{}\xspace}

\def\Bu      {{\ensuremath{\B^+}}\xspace}
\def\Bub     {{\ensuremath{\B^-}}\xspace}
\def\Bp      {{\ensuremath{\Bu}}\xspace}
\def\Bm      {{\ensuremath{\Bub}}\xspace}
\def\Bpm     {{\ensuremath{\B^\pm}}\xspace}

\def\Bd      {{\ensuremath{\B^0}}\xspace}

\def\Bdb     {{\ensuremath{\Bbar{}^0}}\xspace}

  \def\Y#1S{\ensuremath{\PUpsilon{(#1S)}}\xspace}

\def\Lbar        {{\ensuremath{\kern 0.1em\overline{\kern -0.1em\PLambda}}}\xspace}
\def\LorLbar    {\kern 0.18em\optbar{\kern -0.18em \PLambda}{}\xspace}

\newcommand{\decay}[2]{\ensuremath{#1\!\to #2}\xspace}         

\def\to                 {\ensuremath{\rightarrow}\xspace}

\def\order   {{\ensuremath{\mathcal{O}}}\xspace}

\def\CP                {{\ensuremath{C\!P}}\xspace}

\newcommand{\ACP}{{\ensuremath{{\mathcal{A}}^{\CP}}}\xspace}

\def\AT#1     {\ensuremath{A_{\mathrm{T}}^{#1}}\xspace}           

\def\C#1      {\ensuremath{\mathcal{C}_{#1}}\xspace}                       
\def\Cp#1     {\ensuremath{\mathcal{C}_{#1}^{'}}\xspace}                    
\def\Ceff#1   {\ensuremath{\mathcal{C}_{#1}^{\mathrm{(eff)}}}\xspace}        
\def\Cpeff#1  {\ensuremath{\mathcal{C}_{#1}^{'\mathrm{(eff)}}}\xspace}       
\def\Ope#1    {\ensuremath{\mathcal{O}_{#1}}\xspace}                       
\def\Opep#1   {\ensuremath{\mathcal{O}_{#1}^{'}}\xspace}                    

\newcommand{\tev}{\ifthenelse{\boolean{inbibliography}}{\ensuremath{~T\kern -0.05em eV}}{\ensuremath{\mathrm{\,Te\kern -0.1em V}}}\xspace}
\newcommand{\gev}{\ensuremath{\mathrm{\,Ge\kern -0.1em V}}\xspace}
\newcommand{\mev}{\ensuremath{\mathrm{\,Me\kern -0.1em V}}\xspace}
\newcommand{\kev}{\ensuremath{\mathrm{\,ke\kern -0.1em V}}\xspace}
\newcommand{\ev}{\ensuremath{\mathrm{\,e\kern -0.1em V}}\xspace}
\newcommand{\gevc}{\ensuremath{{\mathrm{\,Ge\kern -0.1em V\!/}c}}\xspace}
\newcommand{\mevc}{\ensuremath{{\mathrm{\,Me\kern -0.1em V\!/}c}}\xspace}
\newcommand{\gevcc}{\ensuremath{{\mathrm{\,Ge\kern -0.1em V\!/}c^2}}\xspace}
\newcommand{\gevgevcccc}{\ensuremath{{\mathrm{\,Ge\kern -0.1em V^2\!/}c^4}}\xspace}
\newcommand{\mevcc}{\ensuremath{{\mathrm{\,Me\kern -0.1em V\!/}c^2}}\xspace}

\def\mum  {\ensuremath{{\,\upmu\mathrm{m}}}\xspace}

\def\invfb   {\ensuremath{\mbox{\,fb}^{-1}}\xspace}

\def\order{{\ensuremath{\mathcal{O}}}\xspace}
\newcommand{\chisq}{\ensuremath{\chi^2}\xspace}

\newcommand{\chisqip}{\ensuremath{\chi^2_{\text{IP}}}\xspace}

\def\gsim{{~\raise.15em\hbox{$>$}\kern-.85em
          \lower.35em\hbox{$\sim$}~}\xspace}
\def\lsim{{~\raise.15em\hbox{$<$}\kern-.85em
          \lower.35em\hbox{$\sim$}~}\xspace}

\def\ptot       {\mbox{$p$}\xspace}
\def\pt         {\mbox{$p_{\mathrm{ T}}$}\xspace}

\def\evtgen     {\mbox{\textsc{EvtGen}}\xspace}

\def\geant      {\mbox{\textsc{Geant4}}\xspace}

\def\photos     {\mbox{\textsc{Photos}}\xspace}

\def\pythia     {\mbox{\textsc{Pythia}}\xspace}

\def\tell1  {TELL1\xspace}
\def\ukl1   {UKL1\xspace}

\usepackage{cite} 
\usepackage{mciteplus}

\usepackage{longtable} 
\usepackage[bottom]{footmisc}

\def\DporDs{{\ensuremath{\D^+_{(\squark)}}}\xspace}

\def\DmorDs{{\ensuremath{\D^-_{(\squark)}}}\xspace}
\def\DmorDsstar{{\ensuremath{\D^{*-}_{(\squark)}}}\xspace}
\def\DKpi{{\ensuremath{\decay{\Dz}{K^-\pi^+}}}\xspace}
\def\DKpipipi{{\ensuremath{\decay{\Dz}{K^-\pi^+\pi^-\pi^+}}}\xspace}

\begin{document}

\renewcommand{\thefootnote}{\fnsymbol{footnote}}
\setcounter{footnote}{1}


\begin{titlepage}
\pagenumbering{roman}

\vspace*{-1.5cm}
\centerline{\large EUROPEAN ORGANIZATION FOR NUCLEAR RESEARCH (CERN)}
\vspace*{1.5cm}
\noindent
\begin{tabular*}{\linewidth}{lc@{\extracolsep{\fill}}r@{\extracolsep{0pt}}}
\ifthenelse{\boolean{pdflatex}}
{\vspace*{-1.5cm}\mbox{\!\!\!\includegraphics[width=.14\textwidth]{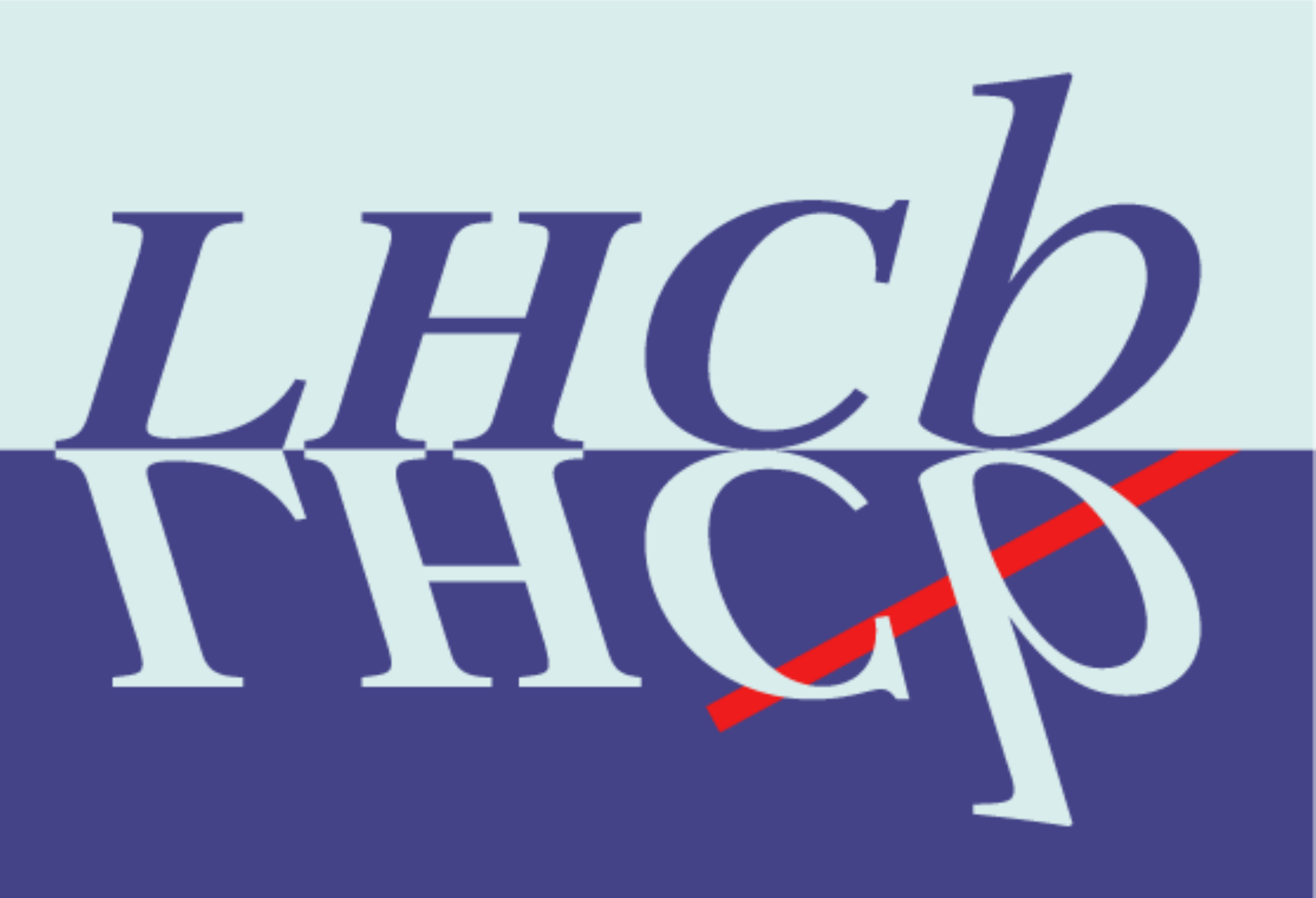}} & &}%
{\vspace*{-1.2cm}\mbox{\!\!\!\includegraphics[width=.12\textwidth]{lhcb-logo.eps}} & &}%
\\
 & & CERN-EP-2018-046 \\  
 & & LHCb-PAPER-2018-007 \\  
 & & 29 March 2018 \\ 
\end{tabular*}

\vspace*{4.0cm}

{\normalfont\bfseries\boldmath\huge
\begin{center}
  \papertitle 
\end{center}
}

\vspace*{2.0cm}

\begin{center}
\paperauthors\footnote{Authors are listed at the end of this paper.}
\end{center}

\vspace{\fill}

\begin{abstract}
  \noindent
The \CP asymmetry in \decay{\Bm}{\Dsm\Dz} and \decay{\Bm}{\Dm\Dz} decays is measured
using LHCb data corresponding to an integrated luminosity of 3.0\invfb, 
collected in $pp$ collisions at centre-of-mass energies of 7 and 8\,TeV.
The results are
${\ACP(\decay{\Bm}{\Dsm\Dz})}=(-0.4\pm 0.5\pm 0.5)\%$ and
${\ACP(\decay{\Bm}{\Dm\Dz})}=( 2.3\pm 2.7\pm 0.4)\%$,
where the first uncertainties are statistical and the second systematic.
This is the first measurement of $\ACP(\decay{\Bm}{\Dsm\Dz})$
and the most precise determination of $\ACP(\decay{\Bm}{\Dm\Dz})$.
Neither result shows evidence of \CP violation.
\end{abstract}

\vspace*{2.0cm}

\begin{center}
  Published in JHEP 05(2018)160
\end{center}

\vspace{\fill}

{\footnotesize 
\centerline{\copyright~\papercopyright, licence \href{\paperlicenceurl}{\paperlicence}.}}
\vspace*{2mm}

\end{titlepage}


\newpage
\setcounter{page}{2}
\mbox{~}

\cleardoublepage


\renewcommand{\thefootnote}{\arabic{footnote}}
\setcounter{footnote}{0}


\pagestyle{plain} 
\setcounter{page}{1}
\pagenumbering{arabic}


%

\setcounter{figure}{0}
\setcounter{table}{0}


\section{Introduction}
\label{sec:Introduction}

Weak decays of heavy hadrons are governed by transition amplitudes that are
proportional to the elements $V_{qq'}$
of the unitary $3 \times 3$  Cabibbo-Kobayashi-Maskawa (CKM) matrix~\cite{Cabibbo:1963yz,Kobayashi:1973fv},
a crucial component of the Standard Model (SM) of elementary particle physics.
Different decay rates between heavy-flavoured hadrons and their antiparticles are possible 
if there is interference between two or more quark-level transitions with different phases.
The corresponding violation of \CP symmetry was first observed in neutral kaon decays~\cite{Christenson:1964fg}.
In \B decays, \CP violation was first observed in the interference between a decay with and without mixing~\cite{Aubert:2001nu,Abe:2001xe}
and later also directly in the decays of \Bd mesons~\cite{Aubert:2004qm,Chao:2004mn}.

\begin{figure}[bp]
\includegraphics[width=8cm]{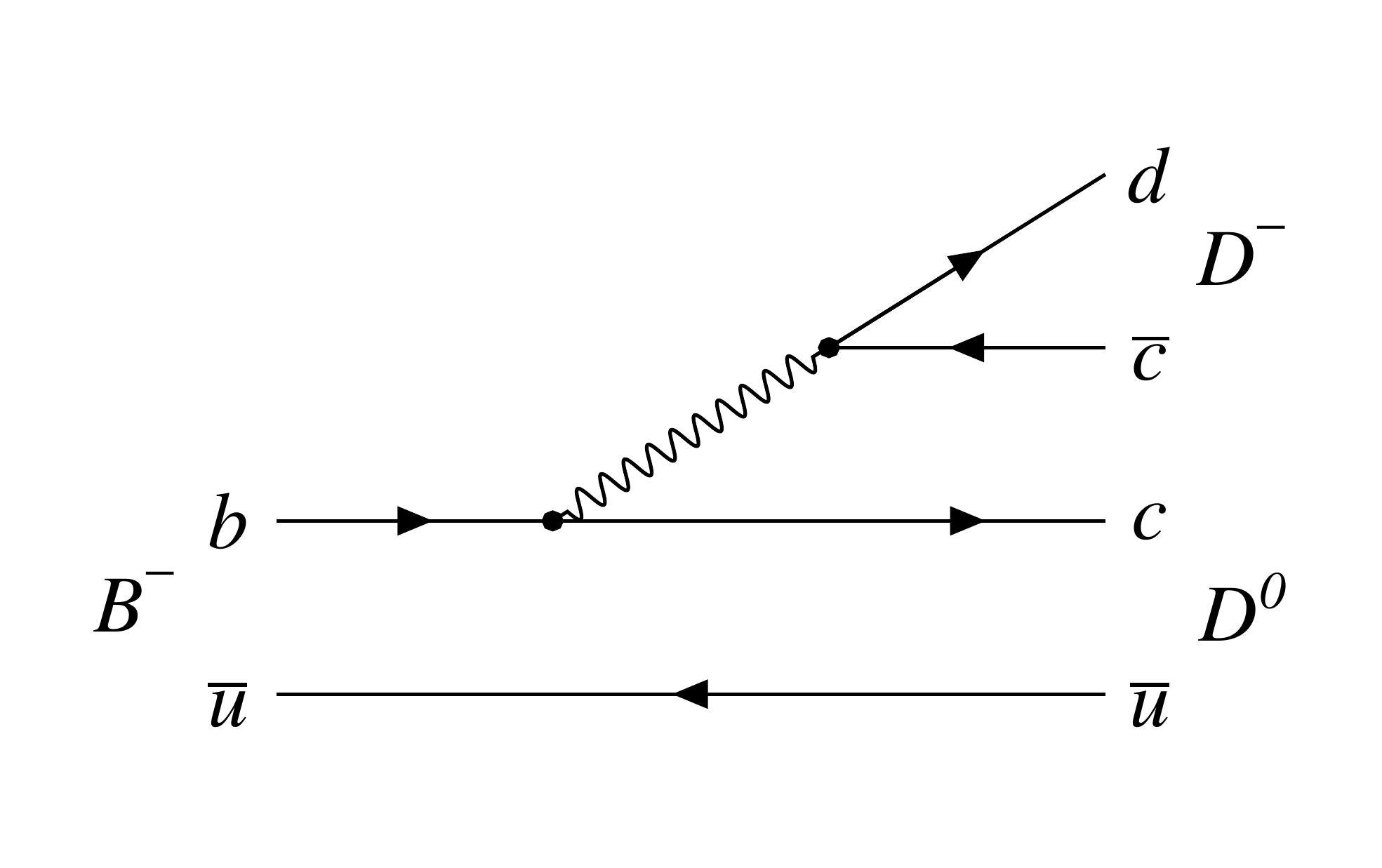}
\includegraphics[width=8cm]{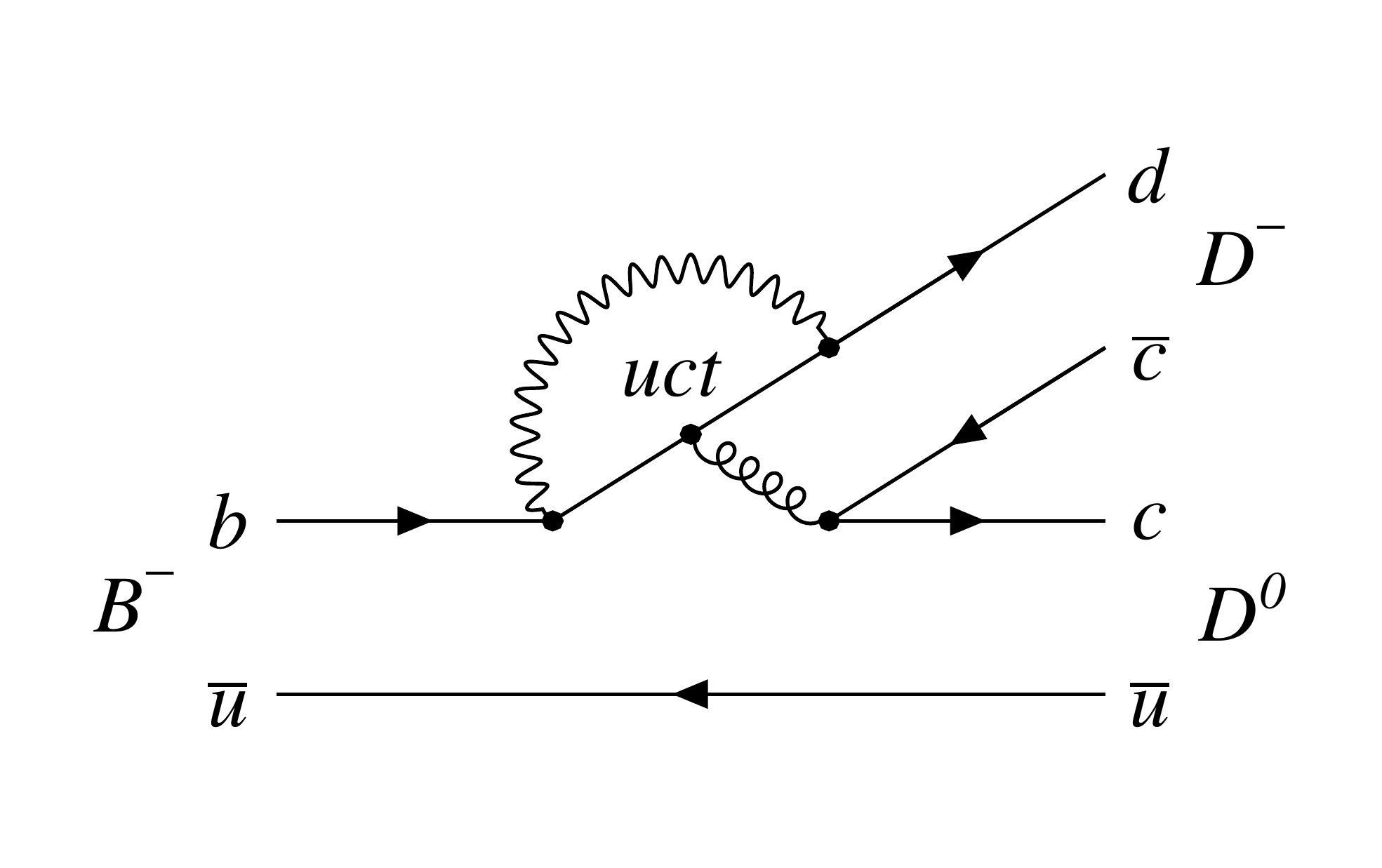}
\caption{Illustration of (left) tree diagram and (right) loop diagram contributions to the decay \decay{\Bm}{\Dm\Dz}.
Similar diagrams, with the \dquark replaced by \squark, apply to the decay \decay{\Bm}{\Dsm\Dz}.}
\label{fig:Diagrams_BpDpDzb}
\end{figure}

The decays of charged or neutral \B mesons to two charm mesons 
are driven by tree-level and loop-level amplitudes, as illustrated in Fig.~\ref{fig:Diagrams_BpDpDzb}.
Annihilation diagrams also contribute, but to a lesser extent.
The decays \decay{\Bdb}{\Dp\Dm}, \decay{\Bdb}{\Dz\Dzb} and \decay{\Bm}{\Dm\Dz}
are related by isospin symmetry,\footnote{Unless
specified otherwise, charge conjugation is implied throughout the paper.}
and expressions that relate the branching fractions and \CP asymmetries,
as well as nonfactorizable effects, 
have been derived~\cite{Sahoo:2017ylz,Bel:2015wha}.

The \CP asymmetry in the decay of the \Bm meson to two charm mesons is defined as
\begin{equation}
\ACP(\decay{\Bm}{\DmorDs\Dz})\equiv\frac{\Gamma(\decay{\Bm}{\DmorDs\Dz})-\Gamma(\decay{\Bp}{\DporDs\Dzb})}{\Gamma(\decay{\Bm}
{\DmorDs\Dz})+\Gamma(\decay{\Bp}{\DporDs\Dzb})}.
\label{eq:ACP}
\end{equation}

Nonzero \CP asymmetries in \decay{\Bm}{\DmorDs\Dz} decays are expected~\cite{Li:2009xf,Fu:2011zzo,Lu:2010gg,Kim:2008ex}
due to interference of contributions from tree-level amplitudes with those from loop-level and annihilation amplitudes.
In the SM, these \CP asymmetries are expected to be small, $\order(10^{-2})$.
New physics contributions can enhance the \CP asymmetry in these decays~\cite{Xu:2016hpp,Lu:2010gg,Jung:2014jfa,Kim:2008ex}.
The most precise measurements of the \CP asymmetry in \decay{\Bm}{\Dm\Dz} decays are from the Belle and BaBar experiments,
$\ACP=(0\pm8\pm2)\%$~\cite{Adachi:2008cj} and
$\ACP=(-13\pm14\pm2)\%$~\cite{Aubert:2006ia}, respectively,
where the first uncertainties are statistical and the second systematic.
The \CP asymmetry in \decay{\Bm}{\Dsm\Dz} decays has not been measured before.

This paper describes a measurement of the \CP asymmetry in \decay{\Bm}{\Dsm\Dz} and 
\decay{\Bm}{\Dm\Dz} decays, using $pp$ collision data corresponding 
to an integrated luminosity of 3.0\invfb, 
of which 1.0\invfb was taken in 2011 at a centre-of-mass energy of
$\sqrt{s}=7$\tev and 2.0\invfb in 2012 at $\sqrt{s}=8$\tev.
Charm mesons are reconstructed in the following decays: 
\mbox{\decay{\Dz}{\Km\pip}},
\mbox{\decay{\Dz}{\Km\pip\pim\pip}},
\mbox{\decay{\Dm}{\Kp\pim\pim}}, and
\mbox{\decay{\Dsm}{\Km\Kp\pim}}.

The determinations of \mbox{$\ACP(\decay{\Bm}{\DmorDs\Dz})$} are based on the measurements of the raw asymmetries
\begin{equation}
A_{\rm raw}\equiv\frac{N(\decay{\Bm}{\DmorDs\Dz})-N(\decay{\Bp}{\DporDs\Dzb})}{N(\decay{\Bm}{\DmorDs\Dz})+N(\decay{\Bp}{\DporDs\Dzb})},
\label{eq:rawACP}
\end{equation}
where $N$ indicates the observed yield in the respective decay channel.
The raw asymmetries include the asymmetry in \B production and detection efficiencies of the final states.
If the asymmetries are small, higher-order terms corresponding to products of the asymmetries can be neglected, and the following relation holds
\begin{equation}
\ACP=A_{\rm raw}-A_P-A_D,
\label{eq:correctACP}
\end{equation}
where $A_P$ is the asymmetry in the production cross-sections, $\sigma$, of \Bpm mesons,
\begin{equation}
A_P\equiv\frac{\sigma(\Bm)-\sigma(\Bp)}{\sigma(\Bm)+\sigma(\Bp)},
\label{eq:A_P}
\end{equation}
and $A_D$ is the asymmetry of the detection efficiencies, $\varepsilon$,
\begin{equation}
A_D\equiv\frac{\varepsilon(\decay{\Bm}{\DmorDs\Dz})-\varepsilon(\decay{\Bp}{\DporDs\Dzb})}{\varepsilon(\decay{\Bm}{\DmorDs\Dz})+\varepsilon(\decay{\Bp}{\DporDs\Dzb})}.
\label{eq:A_D}
\end{equation}


\section{Detector and simulation}
\label{sec:Detector}

The \lhcb detector~\cite{Alves:2008zz,LHCb-DP-2014-002} is a single-arm forward
spectrometer covering the \mbox{pseudorapidity} range $2<\eta <5$,
designed for the study of particles containing \bquark or \cquark
quarks. The detector includes a high-precision tracking system
consisting of a silicon-strip vertex detector surrounding the $pp$
interaction region~\cite{LHCb-DP-2014-001}, a large-area silicon-strip detector located
upstream of a dipole magnet with a bending power of about
$4{\mathrm{\,Tm}}$, and three stations of silicon-strip detectors and straw
drift tubes~\cite{LHCb-DP-2013-003} placed downstream of the magnet.
The polarity of the dipole magnet is reversed periodically throughout data-taking,
to cancel, to first order, asymmetries in the detection efficiency due to  nonuniformities in the detector response.
The configuration with the magnetic field vertically upwards (downwards) bends positively (negatively)
charged particles in the horizontal plane towards the centre of the LHC.

The tracking system provides a measurement of momentum, \ptot, of charged particles with
a relative uncertainty that varies from 0.5\% at low momentum to 1.0\% at 200\gevc.
The minimum distance of a track to a primary vertex (PV), the impact parameter (IP), 
is measured with a resolution of $(15+29/\pt)\mum$,
where \pt is the component of the momentum transverse to the beam, in\,\gevc.
Different types of charged hadrons are distinguished using information
from two ring-imaging Cherenkov (RICH) detectors~\cite{LHCb-DP-2012-003}. 
Photons, electrons and hadrons are identified by a calorimeter system consisting of
scintillating-pad and preshower detectors, an electromagnetic
calorimeter and a hadronic calorimeter. Muons are identified by a
system composed of alternating layers of iron and multiwire
proportional chambers~\cite{LHCb-DP-2012-002}.

The online event selection is performed by a trigger~\cite{LHCb-DP-2012-004}, 
which consists of a hardware stage, based on information from the calorimeter and muon
systems, followed by a software stage, which applies a full event
reconstruction.
 At the hardware trigger stage, events are required to have a muon with high \pt or a
  hadron, photon or electron with high transverse energy in the calorimeters.
The software trigger requires a two-, three- or four-track
  secondary vertex with a large sum of the transverse momenta of
  the tracks and a significant displacement from the primary $pp$
  interaction vertices. At least one track should have $\pt >
  1.7\gevc$ and \chisqip with respect to any
  PV greater than 16, where \chisqip is defined as the
  difference in fit \chisq of a given PV reconstructed with and
  without the considered particle.
 A multivariate algorithm~\cite{BBDT} is used for
  the identification of secondary vertices consistent with the decay
  of a \bquark hadron.

Simulated events are used for the training of a multivariate selection, and for
determining the shape of the invariant-mass distributions of the signals.
In the simulation, $pp$ collisions with \mbox{\decay{\Bm}{\DmorDs\Dz}} decays are generated using
\pythia~\cite{Sjostrand:2007gs,*Sjostrand:2006za} 
 with a specific \lhcb
configuration~\cite{LHCb-PROC-2010-056}.
Decays of hadronic particles
are described by \evtgen~\cite{Lange:2001uf}, in which final-state
radiation is generated using \photos~\cite{Golonka:2005pn}. The
interaction of the generated particles with the detector, and its response,
are implemented using the \geant
toolkit~\cite{Allison:2006ve, *Agostinelli:2002hh} as described in
Ref.~\cite{LHCb-PROC-2011-006}.
Known discrepancies in the simulation for the mass scale, the momentum resolution
and the RICH response are corrected using data-driven methods.

\section{Candidate selection }
\label{sec:selection}

The offline selection of \decay{\Bm}{\DmorDs\Dz} candidates is a two-step process. 
First, loose criteria are applied to select candidates compatible with the decay \decay{\Bm}{\DmorDs\Dz}.
Second, a multivariate selection is applied and optimized by minimizing the statistical
uncertainty on the asymmetry measurement.

Charm meson candidates are constructed by combining 2, 3 or 4 final-state tracks that are incompatible with originating from 
any reconstructed primary vertex ($\chisqip>4$). In addition, the sum of the transverse momenta of the tracks must exceed 1.8\,\gevc,
the invariant mass must be within $\pm25\mevcc$ of the known charm meson mass~\cite{PDG2017} 
and the tracks are required to form a vertex with good fit \chisq.
Particle identification (PID) criteria are also applied to the final-state particles, 
such that particles that have a significantly larger likelihood to be a kaon than a pion 
are not used as a pion candidate, and conversely. 
Three-track combinations that are compatible with both \mbox{\decay{\Dm}{\Kp\pim\pim}} and \mbox{\decay{\Dsm}{\Km\Kp\pim}} decays 
are categorized as either \Dm or \Dsm, 
based on the invariant mass of the three-track combination, 
the compatibility of opposite-charge track combinations with the \mbox{\decay{\phi}{\Kp\Km}} decay, 
and the PID information of the final-state tracks~\cite{LHCb-PAPER-2014-002}.

In events with at least one \Dm or \Dsm candidate and at least one \Dz candidate,
the charm mesons are combined to form a \Bm candidate if their invariant mass is in the range $4.8-7.0\gevcc$.
The \Bm candidate is required to form a vertex with good fit \chisq,
and have a transverse momentum in excess of $4.0\gevc$.
The resulting trajectory of the \Bm candidate must be consistent with originating from the associated PV,
which is the PV for which the \Bm candidate has the smallest value of \chisqip.
The reconstructed decay time divided by its uncertainty, $\tau/\Delta\tau$,
of \Dz and \Dsm mesons with respect to the \Bm vertex is required to exceed $-3$,
while for the longer-lived \Dm meson it is required to exceed $+3$.
The tighter decay-time significance requirement on the \Dm eliminates background
from \mbox{\decay{\Bm}{\Dz\pim\pip\pim}} decays 
where the negatively charged pion is misidentified as a kaon.
In the offline selection, trigger signals are associated with reconstructed particles.
Signal candidates are selected if the trigger decision was due to the candidate itself,
hereafter called trigger on signal (TOS),
or due to the other particles produced in the $pp$ collision,
hereafter called trigger independent of signal (TIS).

The invariant-mass resolution of \decay{\Bm}{\DmorDs\Dz} decays
is significantly improved by performing a constrained fit~\cite{Hulsbergen:2005pu}.
In this fit, the decay products from each vertex are constrained to originate from a common vertex,
the \Bm vertex is constrained to originate from the associated PV, and
the invariant masses of the \Dz and the \DmorDs mesons are constrained to their known masses~\cite{PDG2017},

To reduce the combinatorial background, while keeping the signal efficiency as large as possible,
a multivariate selection based on a boosted decision tree (BDT)~\cite{Breiman,Roe} is applied.
The following variables are used as input to the BDT:
the transverse momentum and the ratio between the likelihoods of the kaon and pion hypotheses of each final-state track;
the fit \chisq of the \Bm candidate and of both charm meson vertices;
the value of \chisqip of the \Bm candidate;
the values of $\tau/\Delta\tau$ for the \Bm and for both charm meson candidates;
the invariant masses of the reconstructed charm meson candidates; and
the invariant masses of opposite-charge tracks from the \DmorDs candidate.
Separate trainings are performed for the \decay{\Bm}{\Dsm\Dz} and the \decay{\Bm}{\Dm}{\Dz} modes, and for both \Dz decay channels.
The BDT is trained using simulated \Bm signal samples
and candidates in the upper mass sideband of the \Bm meson ($5350<m(\DmorDs\Dz)<6200\mevcc$) as background.
To increase the size of the background sample for the BDT training,
the charm meson invariant-mass intervals are increased from $\pm25\mevcc$ to $\pm75\mevcc$,
and `wrong-sign' \decay{\Bm}{\DmorDs\Dzb} candidates are also included.
Checks have been performed to verify that for all the variables used in the BDT 
the simulated \Bm decays describe the observed signals in data well, and that
selections on the BDT output do not alter the shape of the invariant-mass distribution of the
combinatorial background.

The BDT combines all input variables into a single discriminant.
The optimal requirement on this value is determined by maximizing $N_S/\sqrt{N_S+N_B}$,
where $N_S$ is the expected signal yield, determined from the initial signal yield in data multiplied by the BDT efficiency from simulation,
and $N_B$ is the background yield extrapolated from the upper mass sideband to a $\pm20\mevcc$ interval around the \Bm mass.
This selection has an efficiency of 98\% (90\%) for \decay{\Bm}{\Dsm\Dz\ (\Dm\Dz)} decays, 
and a background rejection of 88\% (93\%).

\section{Measurement of the raw asymmetries}
\label{sec:rawasym}

After the event selection, the signal yields and the raw asymmetries are determined by fitting
a model of the invariant-mass distribution of \decay{\Bm}{\DmorDs\Dz} candidates to the data.
The model includes components for the signal decays,
a background from \mbox{\decay{\Bm}{\Km\Kp\pim\Dz}} decays
and a combinatorial background.

The invariant-mass distribution of \decay{\Bm}{\DmorDs}{\Dz} decays is described by a sum of two
Crystal Ball (CB)~\cite{Skwarnicki:1986xj} functions,
with power-law tails proportional to $[m(\DmorDs\Dz)-m(\Bm)]^{-2}$ in opposite directions,
and with a common peak position.
The tail parameters of the CB functions,
as well as the ratio of the widths of both CB components,
are obtained from simulation.
The peak position of the \Bm signal and the width of one of the CB functions
are free parameters in the fits to the data.
This model provides a good description of the \mbox{\decay{\Bm}{\DmorDs\Dz}} signals. 

The Cabibbo-favoured \mbox{\decay{\Bm}{\Km\Kp\pim\Dz}} decay is a background to the \mbox{\decay{\Bm}{\Dsm\Dz}} channel,
despite being strongly suppressed by the invariant-mass requirement on the $\Km\Kp\pim$ mass.
This background is modelled by a single Gaussian function, whose width is determined from a fit to simulated decays and the 
yields determined from the \Dsm sidebands.
The yield of this background is about 30 times smaller than that of the signal, 
and the shape of the invariant-mass distribution is twice as wide.
The combinatorial background is described by an exponential function.
Candidates originating from partially reconstructed \decay{\Bm}{\DmorDsstar\Dz} and \decay{\Bm}{\DmorDs\Dstarz} decays 
do not contribute to the background since their reconstructed invariant mass is below the lower limit of the fit region.

\begin{figure}
\includegraphics[width=7.5cm]{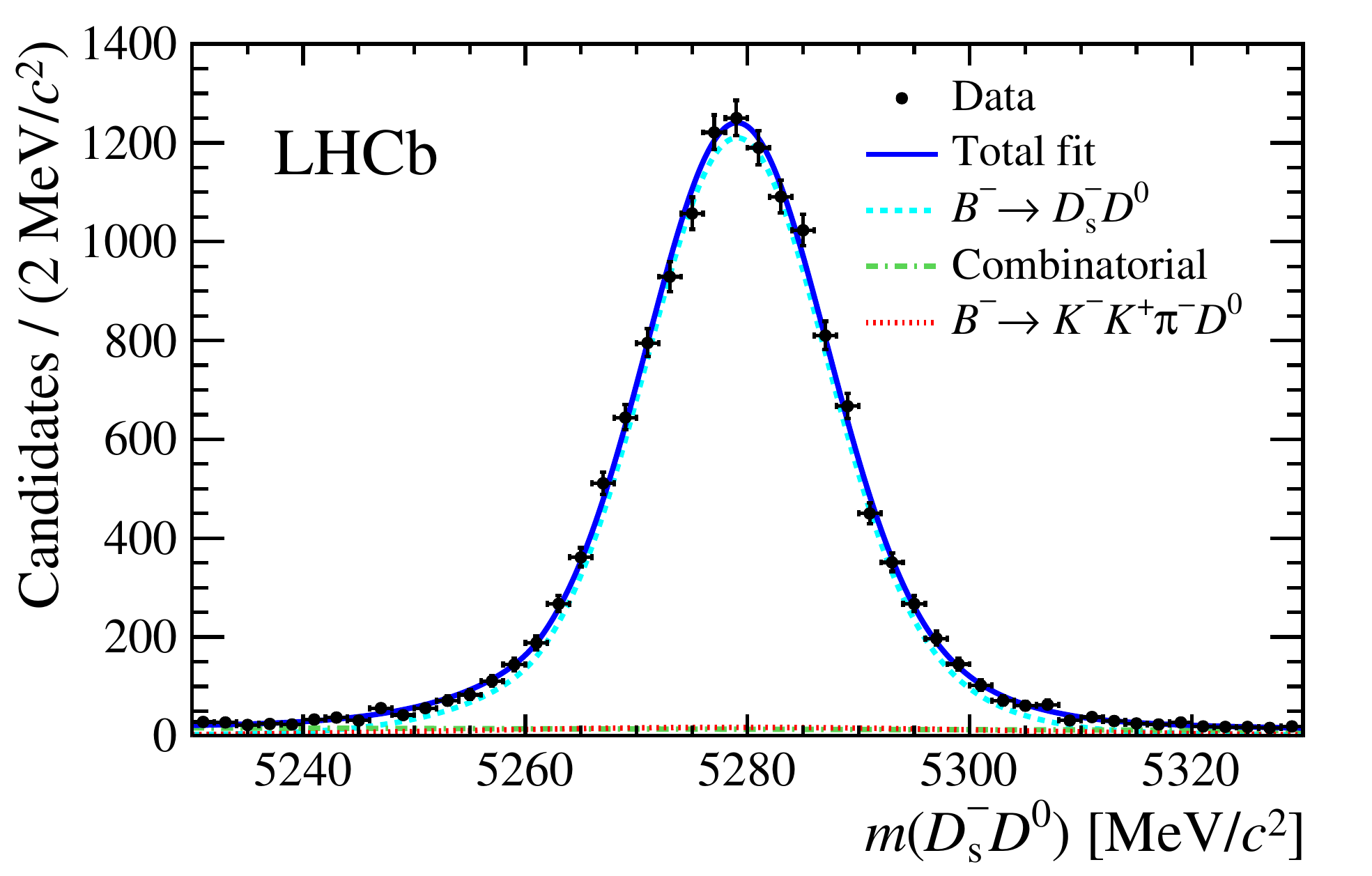}
\includegraphics[width=7.5cm]{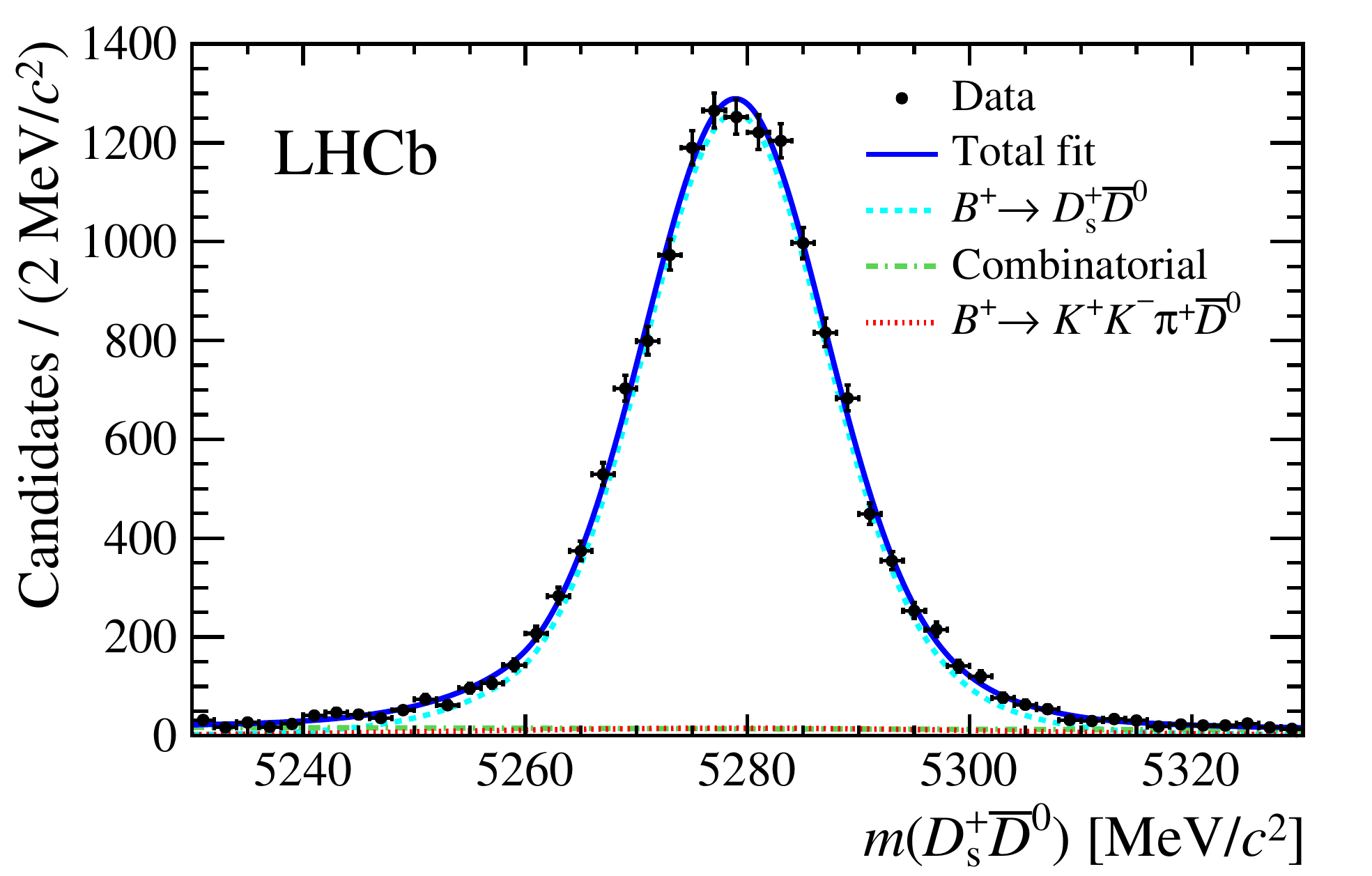}\\
\includegraphics[width=7.5cm]{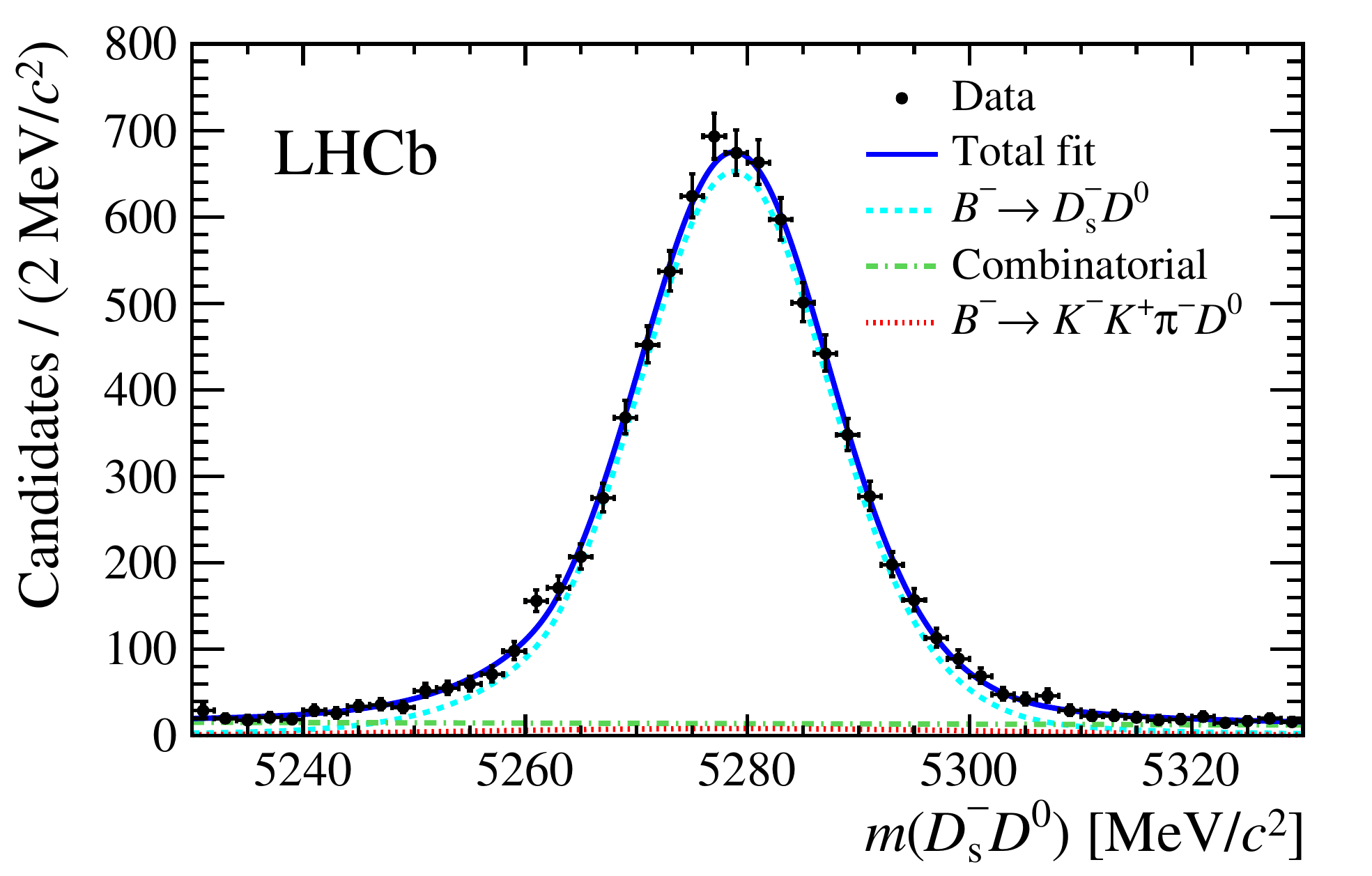}
\includegraphics[width=7.5cm]{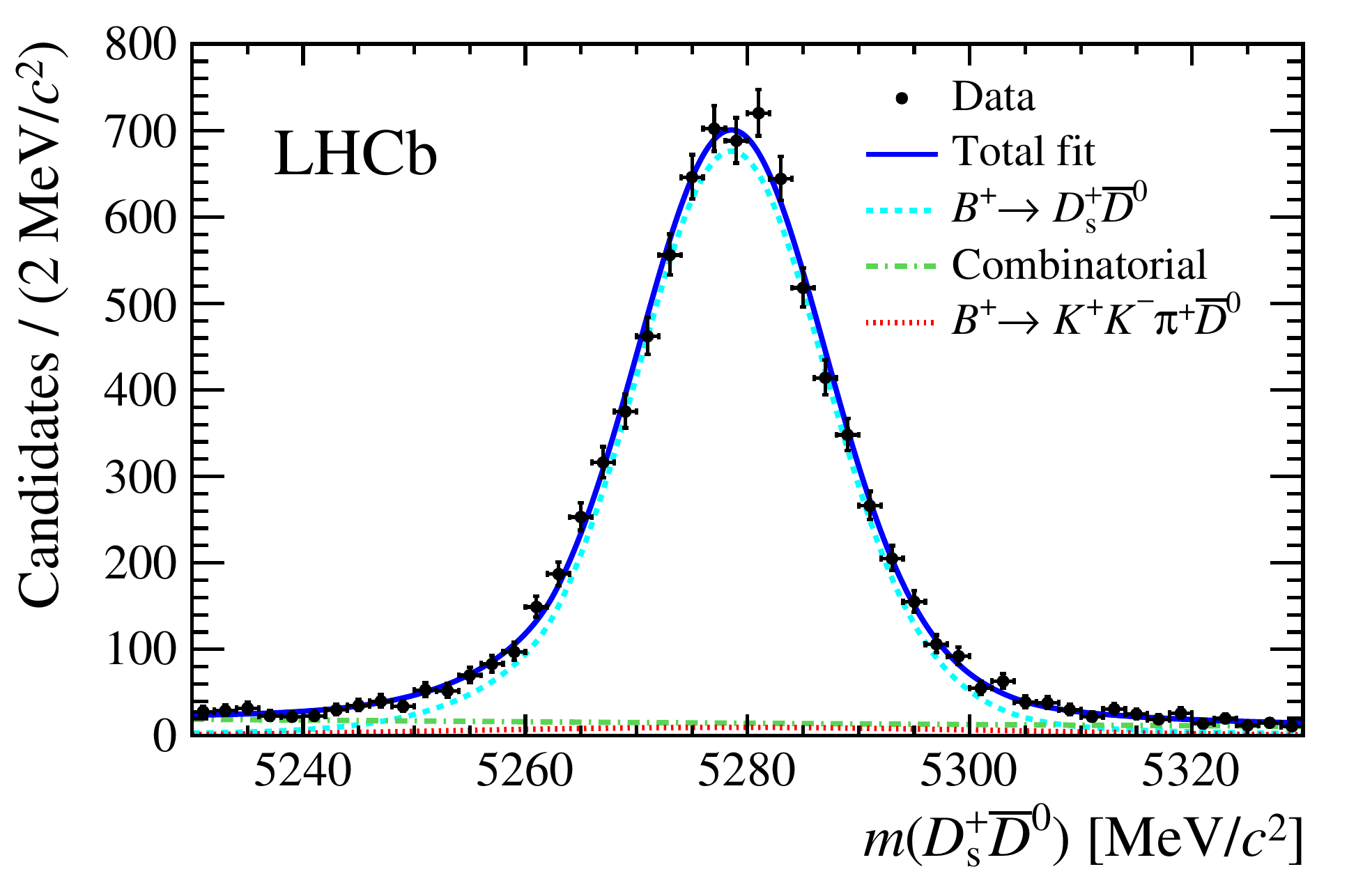}\\
\includegraphics[width=7.5cm]{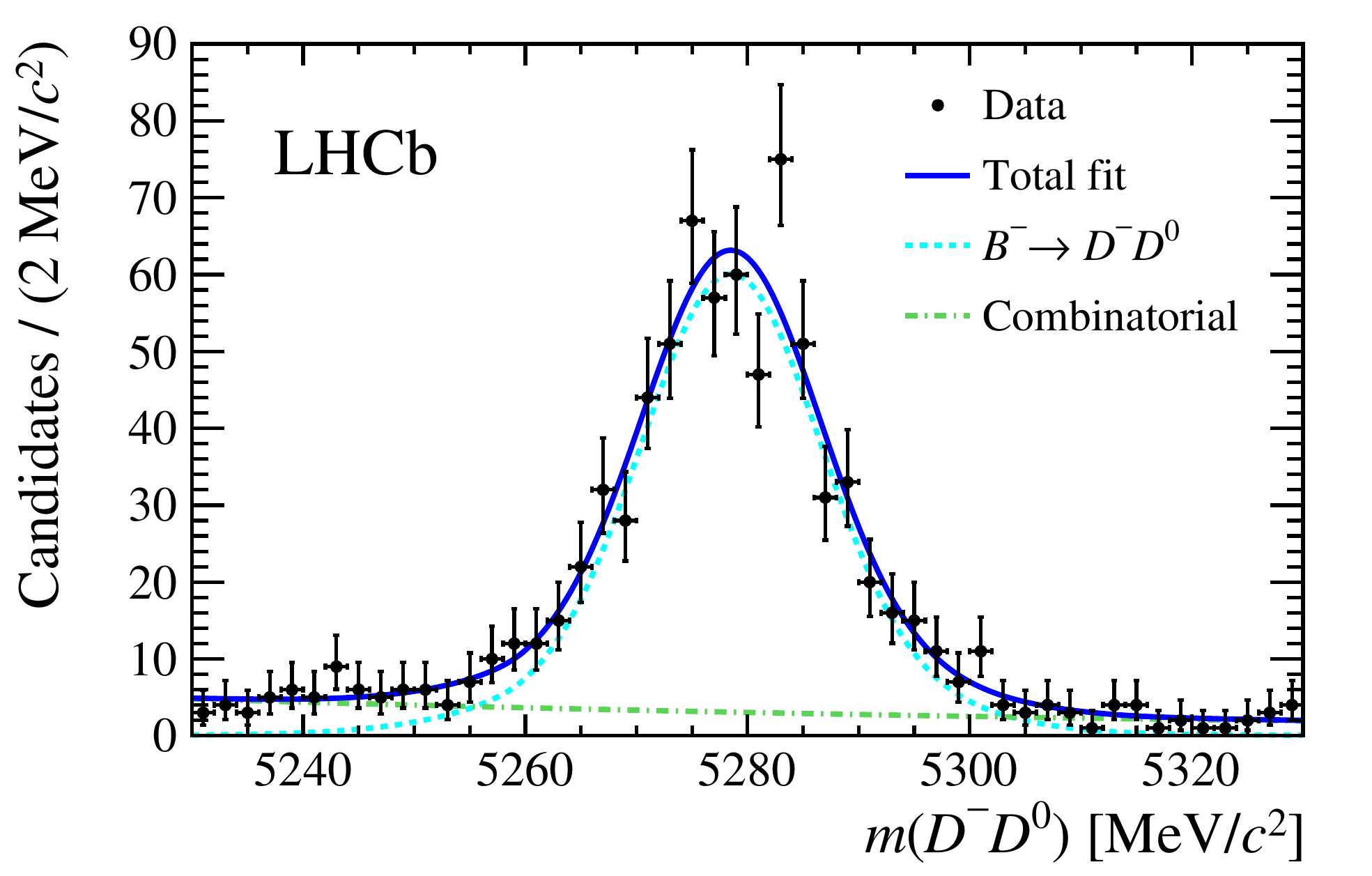}
\includegraphics[width=7.5cm]{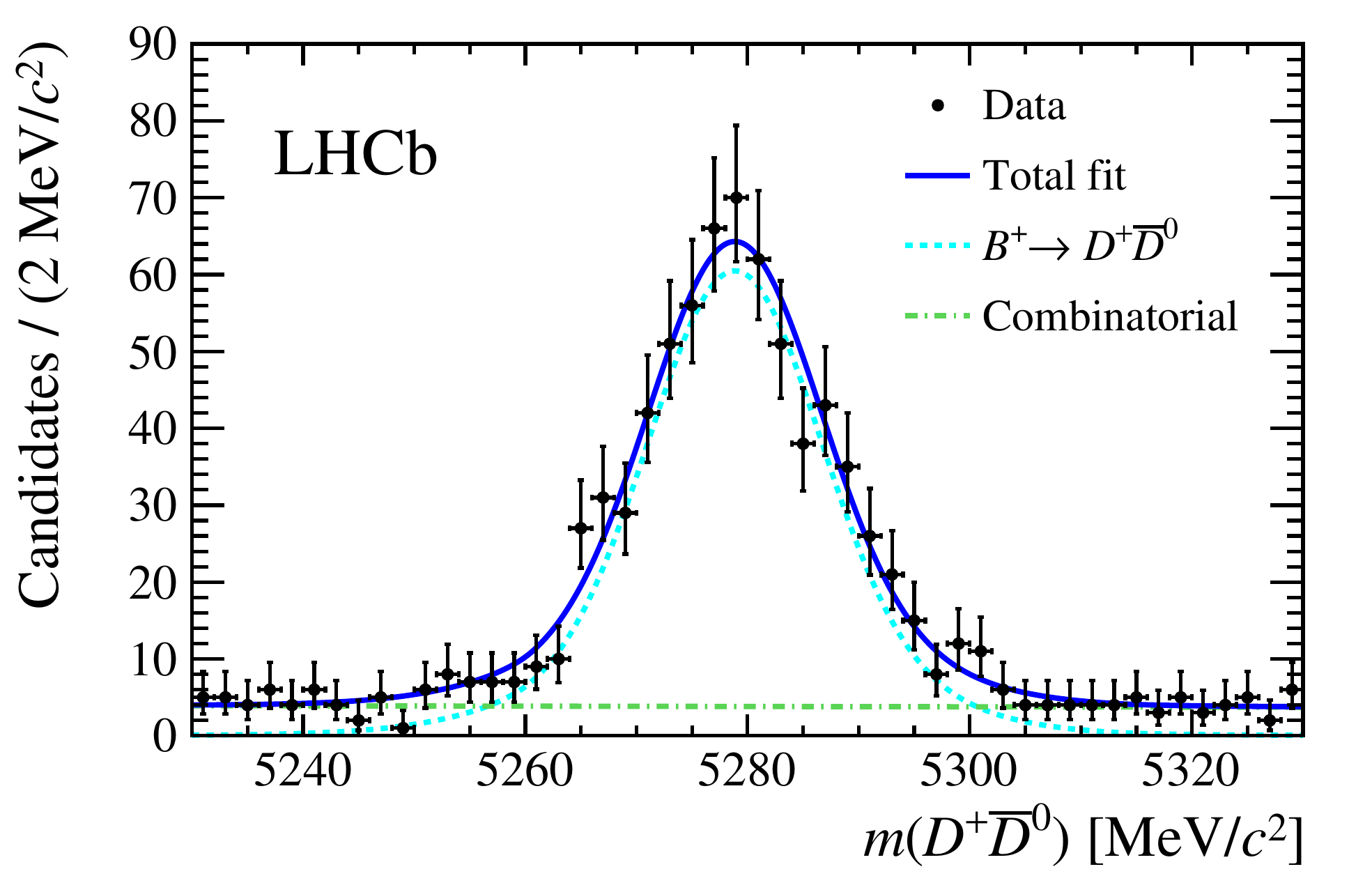}\\
\includegraphics[width=7.5cm]{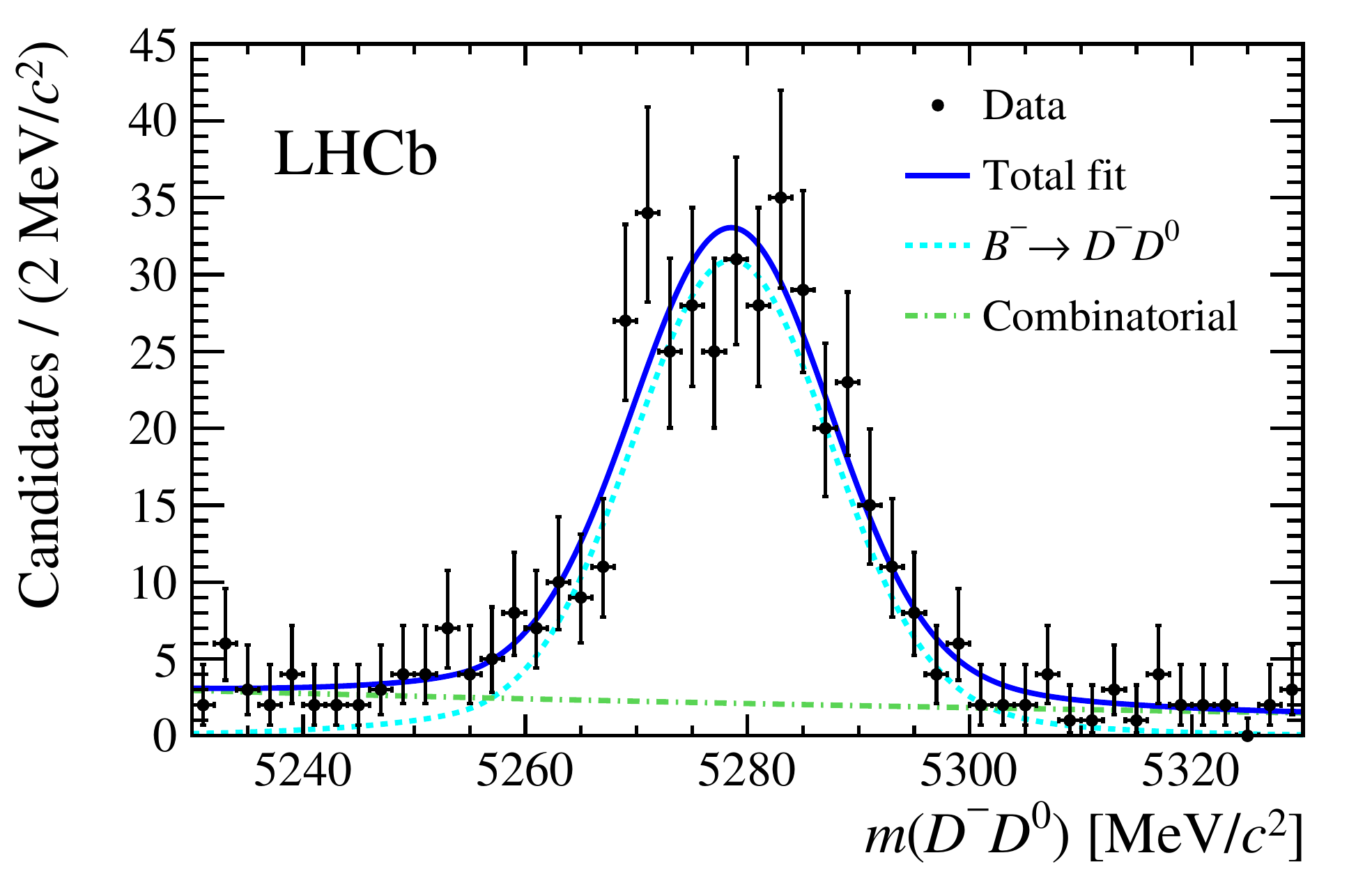}
\includegraphics[width=7.5cm]{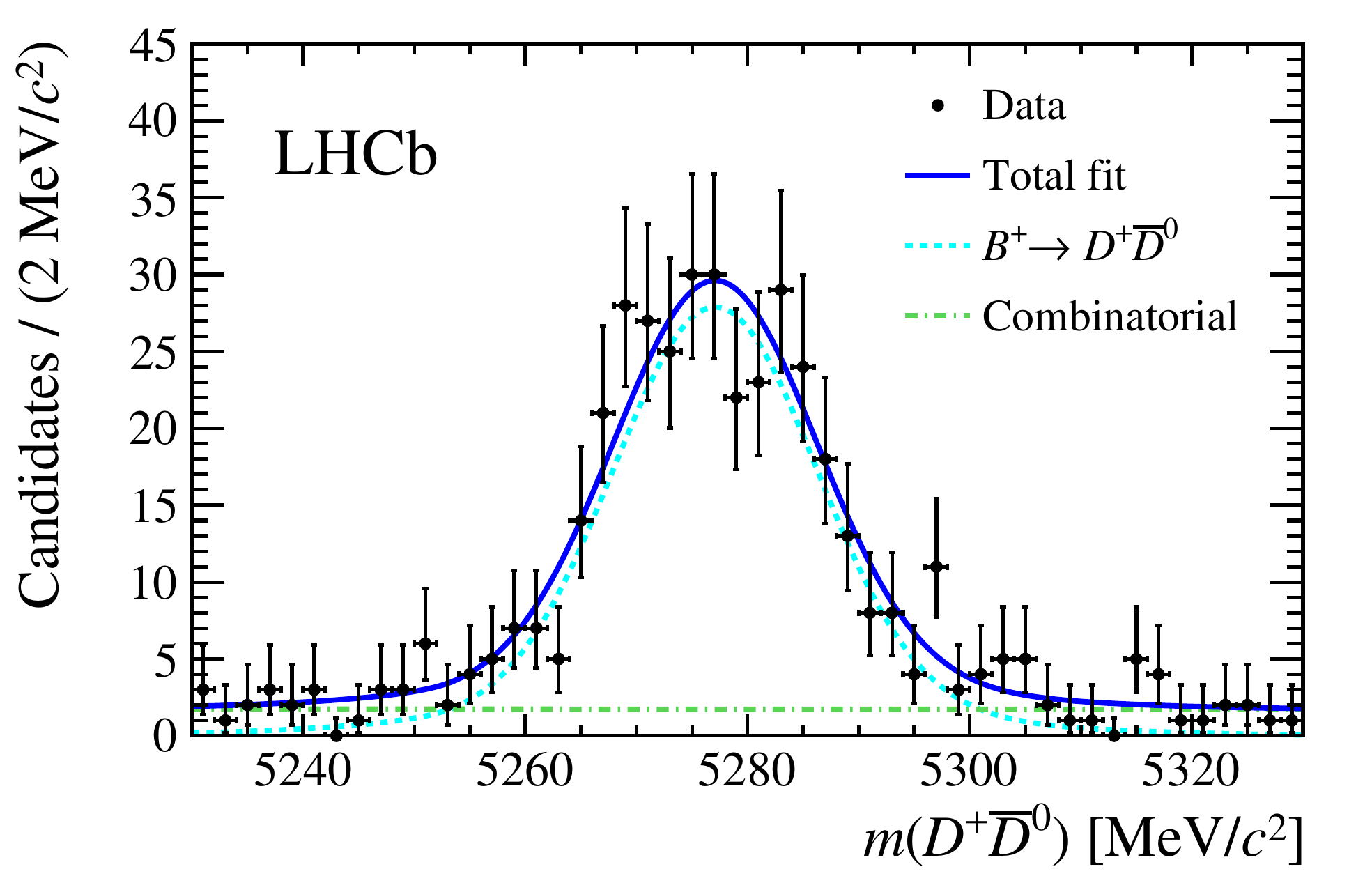}\\
\caption{Invariant-mass distribution of \mbox{\decay{\Bm}{\DmorDs\Dz}} candidates, separated by charge.
The top row plots are \mbox{\decay{\Bm}{\Dsm\Dz}} decays with \DKpi,
the second row with \DKpipipi.
The plots in the third row correspond to  \mbox{\decay{\Bm}{\Dm\Dz}} candidates with \DKpi,
the bottom row with \DKpipipi.
The left plots are \Bm candidates, the right plots \Bp candidates.
The overlaid curves show the fits described in the text.
}
\label{fig:ACPyields}
\end{figure}

Separate unbinned extended maximum likelihood fits are used to describe 
the invariant-mass distributions of candidates with
\mbox{\decay{\Dz}{\Km\pip}} decays and those with \mbox{\decay{\Dz}{\Km\pip\pim\pip}} decays.
Figure~\ref{fig:ACPyields} shows the fits to the invariant-mass
distributions in the fit region, $5230<m(\DmorDs\Dz)<5330\mevcc$, 
of the \mbox{\decay{\Bm}{\Dsm\Dz}} and  \mbox{\decay{\Bm}{\Dm\Dz}} channels, separated by charge and decay mode.
The signal yields and corresponding raw asymmetries, calculated according to Eq.~\ref{eq:rawACP},
are listed in Table~\ref{table:Araw}.
No significant dependence on the magnet polarity or data taking year is observed.
Inaccuracies in the modelling of the signal or background may result in a small biases of the yields,
but are not expected to introduce additional asymmetries, therefore no systematic uncertainties
are attributed to the modelling of the signal and background shapes.

\begin{table}[t]
  \caption{Yields and raw asymmetries for \mbox{\decay{\Bm}{\DmorDs\Dz}} decays.}

  \begin{center}\begin{tabular}{lccc}
  \hline
  Channel                                             & $N(\Bm)$    & $N(\Bp)$ & $A_{\rm raw}$ \\
  \hline
  \decay{\Bm}{\Dsm\Dz}, \DKpi     & $          13659\pm          129$ & $          14209\pm          132$ & $(          -2.0\pm0.7)\%$ \\
  \decay{\Bm}{\Dsm\Dz}, \DKpipipi & $\phantom{0}7717\pm          103$ & $\phantom{0}7945\pm          104$ & $(          -1.5\pm0.9)\%$ \\
  \decay{\Bm}{\Dsm\Dz}, combined  & $          21375\pm          165$ & $          22153\pm          168$ & $(          -1.8\pm0.5)\%$ \\
  \hline                                                              
  \decay{\Bm}{\Dm\Dz}, \DKpi     & $\phantom{00}678\pm\phantom{0}32$ & $\phantom{00}660\pm\phantom{0}31$ & $(\phantom{-}1.3\pm3.3)\%$ \\
  \decay{\Bm}{\Dm\Dz}, \DKpipipi & $\phantom{00}369\pm\phantom{0}24$ & $\phantom{00}345\pm\phantom{0}24$ & $(\phantom{-}3.4\pm4.7)\%$ \\
  \decay{\Bm}{\Dm\Dz}, combined  & $\phantom{0}1047\pm\phantom{0}40$ & $\phantom{0}1005\pm\phantom{0}39$ & $(\phantom{-}2.0\pm2.7)\%$ \\
  \hline
  \end{tabular}\end{center}
\label{table:Araw}
\end{table}

\boldmath
\section{Production and detection asymmetries}
\unboldmath
\label{sec:proddecasym}

The production asymmetry between \Bm and \Bp mesons at LHCb has been measured to be $A_P=(-0.5\pm0.4)\%$
using the \mbox{\decay{\Bm}{\Dz\pim}} decay~\cite{LHCb-PAPER-2016-054}, and no significant
dependence of $A_P$ on the transverse momentum or on the rapidity of the \B meson has been observed.

Four contributions to the asymmetry of the detection efficiencies are considered: 
asymmetries in the tracking efficiency, 
the different $K^\pm$ interaction cross-sections with the detector material, and the trigger and particle identification efficiencies.

The momentum-dependent tracking efficiency for pions has been determined by comparing the yields 
of fully to partially reconstructed \decay{\Dstarp}{(\decay{\Dz}{\Km\pip\pim\pip})\pip} decays~\cite{LHCb-PAPER-2012-009}.
The corresponding asymmetries are summed for all final-state tracks of simulated 
\decay{\Bm}{\DmorDs\Dz} events. After averaging over data-taking year and magnet polarity,
the tracking asymmetry is determined to be $(0.18\pm0.07)\%$ for \decay{\Bm}{\Dsm\Dz} and $(0.21\pm0.07)\%$ for \decay{\Bm}{\Dm\Dz} decays,
where the uncertainties are due to the finite sample of \Dstarp decays used for the tracking efficiency measurement.

The interaction cross-section of \Km mesons with matter is significantly larger than that of \Kp mesons,
resulting in a large asymmetry of the charged kaon detection efficiency.
The momentum-dependent difference in the detection asymmetry between kaons
and pions has been measured by comparing the yield of \mbox{\decay{\Dp}{\Km\pip\pip}} 
to the yield of \mbox{\decay{\Dp}{\KS\pip}} decays~\cite{LHCb-PAPER-2014-013}.
These asymmetries, convoluted with the momentum spectra of the final-state kaons, result in a contribution to the detection asymmetry 
of $(-1.04\pm0.16)\%$ for \decay{\Bm}{\Dsm\Dz} decays,
where the uncertainty is due to the finite samples of \Dp decays.
For \decay{\Bm}{\Dm\Dz} decays, this asymmetry cancels to first order since it has one \Kp and one \Km particle in the final state,
and the resulting asymmetry is $(0.02\pm0.01)\%$. 

The charge asymmetry of TIS candidates is independent of the signal decay channel in consideration and has been measured
in \decay{\Bbar}{\Dz\mun\neumb X} decays~\cite{LHCb-PAPER-2016-054}.
After weighting by the TIS fraction, the asymmetry is found to be 0.04\% and is neglected.
A nonuniform response of the calorimeter may result in a charge asymmetry of the TOS signal.
Large samples of \decay{\Dz}{\Km\pip} decays have been used to determine the \pt-dependent trigger efficiencies and
corresponding charge asymmetries for both pions and kaons.
After convoluting these efficiencies with the simulated \pt spectra, averaging by data-taking year and magnet polarity, 
and multiplying by the TOS fraction of the signal, 
the resulting asymmetry is below 0.05\%, and is considered to be negligible.

In the candidate selection, particle identification criteria that rely on information from the RICH detectors are used.
Possible charge asymmetries in the efficiencies of these selections are studied with samples of \decay{\Dz}{\Km\pip}
that were selected without PID requirements. 
Depending on assumptions on the correlation between the PID and other variables in the multivariate selection, 
asymmetries smaller than 0.1\% are found. Therefore, no correction is applied, and a 0.1\% uncertainty is assigned.

The uncertainties of the contributions to the production and detection asymmetry are considered to be uncorrelated and 
result in a value of $A_P+A_D$ of $(-1.4\pm0.5)\%$ for \decay{\Bm}{\Dsm\Dz} and $(-0.3\pm0.4)\%$ for \decay{\Bm}{\Dm\Dz} decays. 
Changes in the fit model have a negligible effect on the measured asymmetry.

\section{Results and conclusions}
\label{sec:conclusions}

The \CP asymmetries are determined by subtracting the production and detection asymmetries from the measured raw asymmetry according to Eq.~\ref{eq:correctACP}.
The obtained results are
\begin{equation}\nonumber
\ACP(\decay{\Bm}{\Dsm\Dz})=(-0.4\pm0.5\pm0.5)\%,
\end{equation}
\begin{equation}\nonumber
\ACP(\decay{\Bm}{\Dm\Dz})=(\phantom{-}2.3\pm2.7\pm0.4)\%,
\end{equation}
where the first uncertainties are statistical and the second systematic.
The measured value of $\ACP(\decay{\Bm}{\Dsm\Dz})$ provides constraints on the range of 
\CP violation  predicted for a new physics model with $R$-parity violating supersymmetry~\cite{Kim:2008ex}.

In conclusion, the \CP asymmetry in \decay{\Bm}{\Dsm\Dz} decays has been measured for the first time and
the uncertainty on the \CP asymmetry in \decay{\Bm}{\Dm\Dz} decays has been reduced by more
than a factor two with respect to previous measurements.
No evidence for \CP violation in \decay{\Bm}{\DmorDs\Dz} decays has been found.

\section*{Acknowledgements}
%
%
\noindent We express our gratitude to our colleagues in the CERN
accelerator departments for the excellent performance of the LHC. We
thank the technical and administrative staff at the LHCb
institutes. We acknowledge support from CERN and from the national
agencies: CAPES, CNPq, FAPERJ and FINEP (Brazil); MOST and NSFC
(China); CNRS/IN2P3 (France); BMBF, DFG and MPG (Germany); INFN
(Italy); NWO (The Netherlands); MNiSW and NCN (Poland); MEN/IFA
(Romania); MinES and FASO (Russia); MinECo (Spain); SNSF and SER
(Switzerland); NASU (Ukraine); STFC (United Kingdom); NSF (USA).  We
acknowledge the computing resources that are provided by CERN, IN2P3
(France), KIT and DESY (Germany), INFN (Italy), SURF (The
Netherlands), PIC (Spain), GridPP (United Kingdom), RRCKI and Yandex
LLC (Russia), CSCS (Switzerland), IFIN-HH (Romania), CBPF (Brazil),
PL-GRID (Poland) and OSC (USA). We are indebted to the communities
behind the multiple open-source software packages on which we depend.
Individual groups or members have received support from AvH Foundation
(Germany), EPLANET, Marie Sk\l{}odowska-Curie Actions and ERC
(European Union), ANR, Labex P2IO and OCEVU, and R\'{e}gion
Auvergne-Rh\^{o}ne-Alpes (France), Key Research Program of Frontier
Sciences of CAS, CAS PIFI, and the Thousand Talents Program (China),
RFBR, RSF and Yandex LLC (Russia), GVA, XuntaGal and GENCAT (Spain),
Herchel Smith Fund, the Royal Society, the English-Speaking Union and
the Leverhulme Trust (United Kingdom).



\clearpage

\addcontentsline{toc}{section}{References}
\setboolean{inbibliography}{true}
\bibliographystyle{LHCb}
\bibliography{main,LHCb-PAPER,LHCb-CONF,LHCb-DP,LHCb-TDR}

\ifx\mcitethebibliography\mciteundefinedmacro
\PackageError{LHCb.bst}{mciteplus.sty has not been loaded}
{This bibstyle requires the use of the mciteplus package.}\fi
\providecommand{\href}[2]{#2}
\begin{mcitethebibliography}{10}
\mciteSetBstSublistMode{n}
\mciteSetBstMaxWidthForm{subitem}{\alph{mcitesubitemcount})}
\mciteSetBstSublistLabelBeginEnd{\mcitemaxwidthsubitemform\space}
{\relax}{\relax}

\bibitem{Cabibbo:1963yz}
N.~Cabibbo, \ifthenelse{\boolean{articletitles}}{\emph{{Unitary symmetry and
  leptonic decays}},
  }{}\href{http://dx.doi.org/10.1103/PhysRevLett.10.531}{Phys.\ Rev.\ Lett.\
  \textbf{10} (1963) 531}\relax
\mciteBstWouldAddEndPuncttrue
\mciteSetBstMidEndSepPunct{\mcitedefaultmidpunct}
{\mcitedefaultendpunct}{\mcitedefaultseppunct}\relax
\EndOfBibitem
\bibitem{Kobayashi:1973fv}
M.~Kobayashi and T.~Maskawa, \ifthenelse{\boolean{articletitles}}{\emph{{CP
  violation in the renormalizable theory of weak interaction}},
  }{}\href{http://dx.doi.org/10.1143/PTP.49.652}{Prog.\ Theor.\ Phys.\
  \textbf{49} (1973) 652}\relax
\mciteBstWouldAddEndPuncttrue
\mciteSetBstMidEndSepPunct{\mcitedefaultmidpunct}
{\mcitedefaultendpunct}{\mcitedefaultseppunct}\relax
\EndOfBibitem
\bibitem{Christenson:1964fg}
J.~H. Christenson, J.~W. Cronin, V.~L. Fitch, and R.~Turlay,
  \ifthenelse{\boolean{articletitles}}{\emph{{Evidence for the $2\pi$ decay of
  the $K_2^0$ meson}},
  }{}\href{http://dx.doi.org/10.1103/PhysRevLett.13.138}{Phys.\ Rev.\ Lett.\
  \textbf{13} (1964) 138}\relax
\mciteBstWouldAddEndPuncttrue
\mciteSetBstMidEndSepPunct{\mcitedefaultmidpunct}
{\mcitedefaultendpunct}{\mcitedefaultseppunct}\relax
\EndOfBibitem
\bibitem{Aubert:2001nu}
BaBar collaboration, B.~Aubert {\em et~al.},
  \ifthenelse{\boolean{articletitles}}{\emph{{Observation of CP violation in
  the $B^0$ meson system}},
  }{}\href{http://dx.doi.org/10.1103/PhysRevLett.87.091801}{Phys.\ Rev.\ Lett.\
   \textbf{87} (2001) 091801},
  \href{http://arxiv.org/abs/hep-ex/0107013}{{\normalfont\ttfamily
  arXiv:hep-ex/0107013}}\relax
\mciteBstWouldAddEndPuncttrue
\mciteSetBstMidEndSepPunct{\mcitedefaultmidpunct}
{\mcitedefaultendpunct}{\mcitedefaultseppunct}\relax
\EndOfBibitem
\bibitem{Abe:2001xe}
Belle collaboration, K.~Abe {\em et~al.},
  \ifthenelse{\boolean{articletitles}}{\emph{{Observation of large CP violation
  in the neutral $B$ meson system}},
  }{}\href{http://dx.doi.org/10.1103/PhysRevLett.87.091802}{Phys.\ Rev.\ Lett.\
   \textbf{87} (2001) 091802},
  \href{http://arxiv.org/abs/hep-ex/0107061}{{\normalfont\ttfamily
  arXiv:hep-ex/0107061}}\relax
\mciteBstWouldAddEndPuncttrue
\mciteSetBstMidEndSepPunct{\mcitedefaultmidpunct}
{\mcitedefaultendpunct}{\mcitedefaultseppunct}\relax
\EndOfBibitem
\bibitem{Aubert:2004qm}
BaBar collaboration, B.~Aubert {\em et~al.},
  \ifthenelse{\boolean{articletitles}}{\emph{{Direct CP violating asymmetry in
  $B^0 \to K^+ \pi^-$ decays}},
  }{}\href{http://dx.doi.org/10.1103/PhysRevLett.93.131801}{Phys.\ Rev.\ Lett.\
   \textbf{93} (2004) 131801},
  \href{http://arxiv.org/abs/hep-ex/0407057}{{\normalfont\ttfamily
  arXiv:hep-ex/0407057}}\relax
\mciteBstWouldAddEndPuncttrue
\mciteSetBstMidEndSepPunct{\mcitedefaultmidpunct}
{\mcitedefaultendpunct}{\mcitedefaultseppunct}\relax
\EndOfBibitem
\bibitem{Chao:2004mn}
Belle collaboration, Y.~Chao {\em et~al.},
  \ifthenelse{\boolean{articletitles}}{\emph{{Evidence for direct CP violation
  in $B^0 \to K^+ \pi^-$ decays}},
  }{}\href{http://dx.doi.org/10.1103/PhysRevLett.93.191802}{Phys.\ Rev.\ Lett.\
   \textbf{93} (2004) 191802},
  \href{http://arxiv.org/abs/hep-ex/0408100}{{\normalfont\ttfamily
  arXiv:hep-ex/0408100}}\relax
\mciteBstWouldAddEndPuncttrue
\mciteSetBstMidEndSepPunct{\mcitedefaultmidpunct}
{\mcitedefaultendpunct}{\mcitedefaultseppunct}\relax
\EndOfBibitem
\bibitem{Sahoo:2017ylz}
D.~Sahoo {\em et~al.}, \ifthenelse{\boolean{articletitles}}{\emph{{Prediction
  of the CP asymmetry $C_{00}$ in $B^0 \to \Dz\Dzb$ decay}},
  }{}\href{http://dx.doi.org/10.1007/JHEP11(2017)087}{JHEP \textbf{11} (2017)
  087}, \href{http://arxiv.org/abs/1709.08301}{{\normalfont\ttfamily
  arXiv:1709.08301}}\relax
\mciteBstWouldAddEndPuncttrue
\mciteSetBstMidEndSepPunct{\mcitedefaultmidpunct}
{\mcitedefaultendpunct}{\mcitedefaultseppunct}\relax
\EndOfBibitem
\bibitem{Bel:2015wha}
L.~Bel {\em et~al.}, \ifthenelse{\boolean{articletitles}}{\emph{{Anatomy of $
  B\to D\overline{D} $ decays}},
  }{}\href{http://dx.doi.org/10.1007/JHEP07(2015)108}{JHEP \textbf{07} (2015)
  108}, \href{http://arxiv.org/abs/1505.01361}{{\normalfont\ttfamily
  arXiv:1505.01361}}\relax
\mciteBstWouldAddEndPuncttrue
\mciteSetBstMidEndSepPunct{\mcitedefaultmidpunct}
{\mcitedefaultendpunct}{\mcitedefaultseppunct}\relax
\EndOfBibitem
\bibitem{Li:2009xf}
R.-H. Li, X.-X. Wang, A.~I. Sanda, and C.-D. Lu,
  \ifthenelse{\boolean{articletitles}}{\emph{{Decays of $B$ meson to two
  charmed mesons}},
  }{}\href{http://dx.doi.org/10.1103/PhysRevD.81.034006}{Phys.\ Rev.\
  \textbf{D81} (2010) 034006},
  \href{http://arxiv.org/abs/0910.1424}{{\normalfont\ttfamily
  arXiv:0910.1424}}\relax
\mciteBstWouldAddEndPuncttrue
\mciteSetBstMidEndSepPunct{\mcitedefaultmidpunct}
{\mcitedefaultendpunct}{\mcitedefaultseppunct}\relax
\EndOfBibitem
\bibitem{Fu:2011zzo}
H.-F. Fu, G.-L. Wang, Z.-H. Wang, and X.-J. Chen,
  \ifthenelse{\boolean{articletitles}}{\emph{{Semi-leptonic and non-leptonic B
  meson decays to charmed mesons}},
  }{}\href{http://dx.doi.org/10.1088/0256-307X/28/12/121301}{Chin.\ Phys.\
  Lett.\  \textbf{28} (2011) 121301},
  \href{http://arxiv.org/abs/1202.1221}{{\normalfont\ttfamily
  arXiv:1202.1221}}\relax
\mciteBstWouldAddEndPuncttrue
\mciteSetBstMidEndSepPunct{\mcitedefaultmidpunct}
{\mcitedefaultendpunct}{\mcitedefaultseppunct}\relax
\EndOfBibitem
\bibitem{Lu:2010gg}
L.-X. L{\"u}, Z.-J. Xiao, S.-W. Wang, and W.-J. Li,
  \ifthenelse{\boolean{articletitles}}{\emph{{Double charm decays of $B$ mesons
  in the mSUGRA model}},
  }{}\href{http://dx.doi.org/10.1088/0253-6102/56/1/22}{Commun.\ Theor.\ Phys.\
   \textbf{56} (2011) 125},
  \href{http://arxiv.org/abs/1008.4987}{{\normalfont\ttfamily
  arXiv:1008.4987}}\relax
\mciteBstWouldAddEndPuncttrue
\mciteSetBstMidEndSepPunct{\mcitedefaultmidpunct}
{\mcitedefaultendpunct}{\mcitedefaultseppunct}\relax
\EndOfBibitem
\bibitem{Kim:2008ex}
C.~S. Kim, R.-M. Wang, and Y.-D. Yang,
  \ifthenelse{\boolean{articletitles}}{\emph{{Studying double charm decays of
  $B_{u,d}$ and $B_s$ mesons in the MSSM with R-parity violation}},
  }{}\href{http://dx.doi.org/10.1103/PhysRevD.79.055004}{Phys.\ Rev.\
  \textbf{D79} (2009) 055004},
  \href{http://arxiv.org/abs/0812.4136}{{\normalfont\ttfamily
  arXiv:0812.4136}}\relax
\mciteBstWouldAddEndPuncttrue
\mciteSetBstMidEndSepPunct{\mcitedefaultmidpunct}
{\mcitedefaultendpunct}{\mcitedefaultseppunct}\relax
\EndOfBibitem
\bibitem{Xu:2016hpp}
Y.-G. Xu and R.-M. Wang, \ifthenelse{\boolean{articletitles}}{\emph{{Studying
  the fourth generation quark contributions to the double charm decays $B_{(s)}
  \to D_{(s)}^{(*)} D_{s}^{(*)}$}},
  }{}\href{http://dx.doi.org/10.1007/s10773-016-3149-x}{Int.\ J.\ Theor.\
  Phys.\  \textbf{55} (2016) 5290}\relax
\mciteBstWouldAddEndPuncttrue
\mciteSetBstMidEndSepPunct{\mcitedefaultmidpunct}
{\mcitedefaultendpunct}{\mcitedefaultseppunct}\relax
\EndOfBibitem
\bibitem{Jung:2014jfa}
M.~Jung and S.~Schacht, \ifthenelse{\boolean{articletitles}}{\emph{{Standard
  model predictions and new physics sensitivity in $B \to DD$ decays}},
  }{}\href{http://dx.doi.org/10.1103/PhysRevD.91.034027}{Phys.\ Rev.\
  \textbf{D91} (2015) 034027},
  \href{http://arxiv.org/abs/1410.8396}{{\normalfont\ttfamily
  arXiv:1410.8396}}\relax
\mciteBstWouldAddEndPuncttrue
\mciteSetBstMidEndSepPunct{\mcitedefaultmidpunct}
{\mcitedefaultendpunct}{\mcitedefaultseppunct}\relax
\EndOfBibitem
\bibitem{Adachi:2008cj}
Belle collaboration, I.~Adachi {\em et~al.},
  \ifthenelse{\boolean{articletitles}}{\emph{{Measurement of the branching
  fraction and charge asymmetry of the decay \decay{\Bp}{\Dp\Dzb} and search
  for \decay{\Bz}{\Dz\Dzb}}},
  }{}\href{http://dx.doi.org/10.1103/PhysRevD.77.091101}{Phys.\ Rev.\
  \textbf{D77} (2008) 091101},
  \href{http://arxiv.org/abs/0802.2988}{{\normalfont\ttfamily
  arXiv:0802.2988}}\relax
\mciteBstWouldAddEndPuncttrue
\mciteSetBstMidEndSepPunct{\mcitedefaultmidpunct}
{\mcitedefaultendpunct}{\mcitedefaultseppunct}\relax
\EndOfBibitem
\bibitem{Aubert:2006ia}
BaBar collaboration, B.~Aubert {\em et~al.},
  \ifthenelse{\boolean{articletitles}}{\emph{{Measurement of branching
  fractions and CP-violating charge asymmetries for \B-meson decays to
  $\D^{(*)}\Dbar{}^{(*)}$, and implications for the Cabibbo-Kobayashi-Maskawa
  angle $\gamma$}},
  }{}\href{http://dx.doi.org/10.1103/PhysRevD.73.112004}{Phys.\ Rev.\
  \textbf{D73} (2006) 112004},
  \href{http://arxiv.org/abs/hep-ex/0604037}{{\normalfont\ttfamily
  arXiv:hep-ex/0604037}}\relax
\mciteBstWouldAddEndPuncttrue
\mciteSetBstMidEndSepPunct{\mcitedefaultmidpunct}
{\mcitedefaultendpunct}{\mcitedefaultseppunct}\relax
\EndOfBibitem
\bibitem{Alves:2008zz}
LHCb collaboration, A.~A. Alves~Jr.\ {\em et~al.},
  \ifthenelse{\boolean{articletitles}}{\emph{{The \lhcb detector at the LHC}},
  }{}\href{http://dx.doi.org/10.1088/1748-0221/3/08/S08005}{JINST \textbf{3}
  (2008) S08005}\relax
\mciteBstWouldAddEndPuncttrue
\mciteSetBstMidEndSepPunct{\mcitedefaultmidpunct}
{\mcitedefaultendpunct}{\mcitedefaultseppunct}\relax
\EndOfBibitem
\bibitem{LHCb-DP-2014-002}
LHCb collaboration, R.~Aaij {\em et~al.},
  \ifthenelse{\boolean{articletitles}}{\emph{{LHCb detector performance}},
  }{}\href{http://dx.doi.org/10.1142/S0217751X15300227}{Int.\ J.\ Mod.\ Phys.\
  \textbf{A30} (2015) 1530022},
  \href{http://arxiv.org/abs/1412.6352}{{\normalfont\ttfamily
  arXiv:1412.6352}}\relax
\mciteBstWouldAddEndPuncttrue
\mciteSetBstMidEndSepPunct{\mcitedefaultmidpunct}
{\mcitedefaultendpunct}{\mcitedefaultseppunct}\relax
\EndOfBibitem
\bibitem{LHCb-DP-2014-001}
R.~Aaij {\em et~al.}, \ifthenelse{\boolean{articletitles}}{\emph{{Performance
  of the LHCb Vertex Locator}},
  }{}\href{http://dx.doi.org/10.1088/1748-0221/9/09/P09007}{JINST \textbf{9}
  (2014) P09007}, \href{http://arxiv.org/abs/1405.7808}{{\normalfont\ttfamily
  arXiv:1405.7808}}\relax
\mciteBstWouldAddEndPuncttrue
\mciteSetBstMidEndSepPunct{\mcitedefaultmidpunct}
{\mcitedefaultendpunct}{\mcitedefaultseppunct}\relax
\EndOfBibitem
\bibitem{LHCb-DP-2013-003}
R.~Arink {\em et~al.}, \ifthenelse{\boolean{articletitles}}{\emph{{Performance
  of the LHCb Outer Tracker}},
  }{}\href{http://dx.doi.org/10.1088/1748-0221/9/01/P01002}{JINST \textbf{9}
  (2014) P01002}, \href{http://arxiv.org/abs/1311.3893}{{\normalfont\ttfamily
  arXiv:1311.3893}}\relax
\mciteBstWouldAddEndPuncttrue
\mciteSetBstMidEndSepPunct{\mcitedefaultmidpunct}
{\mcitedefaultendpunct}{\mcitedefaultseppunct}\relax
\EndOfBibitem
\bibitem{LHCb-DP-2012-003}
M.~Adinolfi {\em et~al.},
  \ifthenelse{\boolean{articletitles}}{\emph{{Performance of the \lhcb RICH
  detector at the LHC}},
  }{}\href{http://dx.doi.org/10.1140/epjc/s10052-013-2431-9}{Eur.\ Phys.\ J.\
  \textbf{C73} (2013) 2431},
  \href{http://arxiv.org/abs/1211.6759}{{\normalfont\ttfamily
  arXiv:1211.6759}}\relax
\mciteBstWouldAddEndPuncttrue
\mciteSetBstMidEndSepPunct{\mcitedefaultmidpunct}
{\mcitedefaultendpunct}{\mcitedefaultseppunct}\relax
\EndOfBibitem
\bibitem{LHCb-DP-2012-002}
A.~A. Alves~Jr.\ {\em et~al.},
  \ifthenelse{\boolean{articletitles}}{\emph{{Performance of the LHCb muon
  system}}, }{}\href{http://dx.doi.org/10.1088/1748-0221/8/02/P02022}{JINST
  \textbf{8} (2013) P02022},
  \href{http://arxiv.org/abs/1211.1346}{{\normalfont\ttfamily
  arXiv:1211.1346}}\relax
\mciteBstWouldAddEndPuncttrue
\mciteSetBstMidEndSepPunct{\mcitedefaultmidpunct}
{\mcitedefaultendpunct}{\mcitedefaultseppunct}\relax
\EndOfBibitem
\bibitem{LHCb-DP-2012-004}
R.~Aaij {\em et~al.}, \ifthenelse{\boolean{articletitles}}{\emph{{The \lhcb
  trigger and its performance in 2011}},
  }{}\href{http://dx.doi.org/10.1088/1748-0221/8/04/P04022}{JINST \textbf{8}
  (2013) P04022}, \href{http://arxiv.org/abs/1211.3055}{{\normalfont\ttfamily
  arXiv:1211.3055}}\relax
\mciteBstWouldAddEndPuncttrue
\mciteSetBstMidEndSepPunct{\mcitedefaultmidpunct}
{\mcitedefaultendpunct}{\mcitedefaultseppunct}\relax
\EndOfBibitem
\bibitem{BBDT}
V.~V. Gligorov and M.~Williams,
  \ifthenelse{\boolean{articletitles}}{\emph{{Efficient, reliable and fast
  high-level triggering using a bonsai boosted decision tree}},
  }{}\href{http://dx.doi.org/10.1088/1748-0221/8/02/P02013}{JINST \textbf{8}
  (2013) P02013}, \href{http://arxiv.org/abs/1210.6861}{{\normalfont\ttfamily
  arXiv:1210.6861}}\relax
\mciteBstWouldAddEndPuncttrue
\mciteSetBstMidEndSepPunct{\mcitedefaultmidpunct}
{\mcitedefaultendpunct}{\mcitedefaultseppunct}\relax
\EndOfBibitem
\bibitem{Sjostrand:2007gs}
T.~Sj\"{o}strand, S.~Mrenna, and P.~Skands,
  \ifthenelse{\boolean{articletitles}}{\emph{{A brief introduction to PYTHIA
  8.1}}, }{}\href{http://dx.doi.org/10.1016/j.cpc.2008.01.036}{Comput.\ Phys.\
  Commun.\  \textbf{178} (2008) 852},
  \href{http://arxiv.org/abs/0710.3820}{{\normalfont\ttfamily
  arXiv:0710.3820}}\relax
\mciteBstWouldAddEndPuncttrue
\mciteSetBstMidEndSepPunct{\mcitedefaultmidpunct}
{\mcitedefaultendpunct}{\mcitedefaultseppunct}\relax
\EndOfBibitem
\bibitem{Sjostrand:2006za}
T.~Sj\"{o}strand, S.~Mrenna, and P.~Skands,
  \ifthenelse{\boolean{articletitles}}{\emph{{PYTHIA 6.4 physics and manual}},
  }{}\href{http://dx.doi.org/10.1088/1126-6708/2006/05/026}{JHEP \textbf{05}
  (2006) 026}, \href{http://arxiv.org/abs/hep-ph/0603175}{{\normalfont\ttfamily
  arXiv:hep-ph/0603175}}\relax
\mciteBstWouldAddEndPuncttrue
\mciteSetBstMidEndSepPunct{\mcitedefaultmidpunct}
{\mcitedefaultendpunct}{\mcitedefaultseppunct}\relax
\EndOfBibitem
\bibitem{LHCb-PROC-2010-056}
I.~Belyaev {\em et~al.}, \ifthenelse{\boolean{articletitles}}{\emph{{Handling
  of the generation of primary events in Gauss, the LHCb simulation
  framework}}, }{}\href{http://dx.doi.org/10.1088/1742-6596/331/3/032047}{{J.\
  Phys.\ Conf.\ Ser.\ } \textbf{331} (2011) 032047}\relax
\mciteBstWouldAddEndPuncttrue
\mciteSetBstMidEndSepPunct{\mcitedefaultmidpunct}
{\mcitedefaultendpunct}{\mcitedefaultseppunct}\relax
\EndOfBibitem
\bibitem{Lange:2001uf}
D.~J. Lange, \ifthenelse{\boolean{articletitles}}{\emph{{The EvtGen particle
  decay simulation package}},
  }{}\href{http://dx.doi.org/10.1016/S0168-9002(01)00089-4}{Nucl.\ Instrum.\
  Meth.\  \textbf{A462} (2001) 152}\relax
\mciteBstWouldAddEndPuncttrue
\mciteSetBstMidEndSepPunct{\mcitedefaultmidpunct}
{\mcitedefaultendpunct}{\mcitedefaultseppunct}\relax
\EndOfBibitem
\bibitem{Golonka:2005pn}
P.~Golonka and Z.~Was, \ifthenelse{\boolean{articletitles}}{\emph{{PHOTOS Monte
  Carlo: A precision tool for QED corrections in $Z$ and $W$ decays}},
  }{}\href{http://dx.doi.org/10.1140/epjc/s2005-02396-4}{Eur.\ Phys.\ J.\
  \textbf{C45} (2006) 97},
  \href{http://arxiv.org/abs/hep-ph/0506026}{{\normalfont\ttfamily
  arXiv:hep-ph/0506026}}\relax
\mciteBstWouldAddEndPuncttrue
\mciteSetBstMidEndSepPunct{\mcitedefaultmidpunct}
{\mcitedefaultendpunct}{\mcitedefaultseppunct}\relax
\EndOfBibitem
\bibitem{Allison:2006ve}
Geant4 collaboration, J.~Allison {\em et~al.},
  \ifthenelse{\boolean{articletitles}}{\emph{{Geant4 developments and
  applications}}, }{}\href{http://dx.doi.org/10.1109/TNS.2006.869826}{IEEE
  Trans.\ Nucl.\ Sci.\  \textbf{53} (2006) 270}\relax
\mciteBstWouldAddEndPuncttrue
\mciteSetBstMidEndSepPunct{\mcitedefaultmidpunct}
{\mcitedefaultendpunct}{\mcitedefaultseppunct}\relax
\EndOfBibitem
\bibitem{Agostinelli:2002hh}
Geant4 collaboration, S.~Agostinelli {\em et~al.},
  \ifthenelse{\boolean{articletitles}}{\emph{{Geant4: A simulation toolkit}},
  }{}\href{http://dx.doi.org/10.1016/S0168-9002(03)01368-8}{Nucl.\ Instrum.\
  Meth.\  \textbf{A506} (2003) 250}\relax
\mciteBstWouldAddEndPuncttrue
\mciteSetBstMidEndSepPunct{\mcitedefaultmidpunct}
{\mcitedefaultendpunct}{\mcitedefaultseppunct}\relax
\EndOfBibitem
\bibitem{LHCb-PROC-2011-006}
M.~Clemencic {\em et~al.}, \ifthenelse{\boolean{articletitles}}{\emph{{The
  \lhcb simulation application, Gauss: Design, evolution and experience}},
  }{}\href{http://dx.doi.org/10.1088/1742-6596/331/3/032023}{{J.\ Phys.\ Conf.\
  Ser.\ } \textbf{331} (2011) 032023}\relax
\mciteBstWouldAddEndPuncttrue
\mciteSetBstMidEndSepPunct{\mcitedefaultmidpunct}
{\mcitedefaultendpunct}{\mcitedefaultseppunct}\relax
\EndOfBibitem
\bibitem{PDG2017}
Particle Data Group, C.~Patrignani {\em et~al.},
  \ifthenelse{\boolean{articletitles}}{\emph{{\href{http://pdg.lbl.gov/}{Review
  of particle physics}}},
  }{}\href{http://dx.doi.org/10.1088/1674-1137/40/10/100001}{Chin.\ Phys.\
  \textbf{C40} (2016) 100001}, {and 2017 update}\relax
\mciteBstWouldAddEndPuncttrue
\mciteSetBstMidEndSepPunct{\mcitedefaultmidpunct}
{\mcitedefaultendpunct}{\mcitedefaultseppunct}\relax
\EndOfBibitem
\bibitem{LHCb-PAPER-2014-002}
LHCb collaboration, R.~Aaij {\em et~al.},
  \ifthenelse{\boolean{articletitles}}{\emph{{Study of beauty hadron decays
  into pairs of charm hadrons}},
  }{}\href{http://dx.doi.org/10.1103/PhysRevLett.112.202001}{Phys.\ Rev.\
  Lett.\  \textbf{112} (2014) 202001},
  \href{http://arxiv.org/abs/1403.3606}{{\normalfont\ttfamily
  arXiv:1403.3606}}\relax
\mciteBstWouldAddEndPuncttrue
\mciteSetBstMidEndSepPunct{\mcitedefaultmidpunct}
{\mcitedefaultendpunct}{\mcitedefaultseppunct}\relax
\EndOfBibitem
\bibitem{Hulsbergen:2005pu}
W.~D. Hulsbergen, \ifthenelse{\boolean{articletitles}}{\emph{{Decay chain
  fitting with a Kalman filter}},
  }{}\href{http://dx.doi.org/10.1016/j.nima.2005.06.078}{Nucl.\ Instrum.\
  Meth.\  \textbf{A552} (2005) 566},
  \href{http://arxiv.org/abs/physics/0503191}{{\normalfont\ttfamily
  arXiv:physics/0503191}}\relax
\mciteBstWouldAddEndPuncttrue
\mciteSetBstMidEndSepPunct{\mcitedefaultmidpunct}
{\mcitedefaultendpunct}{\mcitedefaultseppunct}\relax
\EndOfBibitem
\bibitem{Breiman}
L.~Breiman, J.~H. Friedman, R.~A. Olshen, and C.~J. Stone, {\em Classification
  and regression trees}, Wadsworth international group, Belmont, California,
  USA, 1984\relax
\mciteBstWouldAddEndPuncttrue
\mciteSetBstMidEndSepPunct{\mcitedefaultmidpunct}
{\mcitedefaultendpunct}{\mcitedefaultseppunct}\relax
\EndOfBibitem
\bibitem{Roe}
B.~P. Roe {\em et~al.}, \ifthenelse{\boolean{articletitles}}{\emph{{Boosted
  decision trees as an alternative to artificial neural networks for particle
  identification}},
  }{}\href{http://dx.doi.org/10.1016/j.nima.2004.12.018}{Nucl.\ Instrum.\
  Meth.\  \textbf{A543} (2005) 577},
  \href{http://arxiv.org/abs/physics/0408124}{{\normalfont\ttfamily
  arXiv:physics/0408124}}\relax
\mciteBstWouldAddEndPuncttrue
\mciteSetBstMidEndSepPunct{\mcitedefaultmidpunct}
{\mcitedefaultendpunct}{\mcitedefaultseppunct}\relax
\EndOfBibitem
\bibitem{Skwarnicki:1986xj}
T.~Skwarnicki, {\em {A study of the radiative cascade transitions between the
  Upsilon-prime and Upsilon resonances}}, PhD thesis, Institute of Nuclear
  Physics, Krakow, 1986,
  {\href{http://inspirehep.net/record/230779/}{DESY-F31-86-02}}\relax
\mciteBstWouldAddEndPuncttrue
\mciteSetBstMidEndSepPunct{\mcitedefaultmidpunct}
{\mcitedefaultendpunct}{\mcitedefaultseppunct}\relax
\EndOfBibitem
\bibitem{LHCb-PAPER-2016-054}
LHCb collaboration, R.~Aaij {\em et~al.},
  \ifthenelse{\boolean{articletitles}}{\emph{{Measurement of the $B^\pm$
  production asymmetry and the $C\!P$ asymmetry in $B^\pm\to \jpsi K^\pm$
  decays}}, }{}\href{http://dx.doi.org/10.1103/PhysRevD.95.052005}{Phys.\ Rev.\
   \textbf{D95} (2017) 052005},
  \href{http://arxiv.org/abs/1701.05501}{{\normalfont\ttfamily
  arXiv:1701.05501}}\relax
\mciteBstWouldAddEndPuncttrue
\mciteSetBstMidEndSepPunct{\mcitedefaultmidpunct}
{\mcitedefaultendpunct}{\mcitedefaultseppunct}\relax
\EndOfBibitem
\bibitem{LHCb-PAPER-2012-009}
LHCb collaboration, R.~Aaij {\em et~al.},
  \ifthenelse{\boolean{articletitles}}{\emph{{Measurement of the $\Dsp$--$\Dsm$
  production asymmetry in $7$\tev $\proton\proton$ collisions}},
  }{}\href{http://dx.doi.org/10.1016/j.physletb.2012.06.001}{Phys.\ Lett.\
  \textbf{B713} (2012) 186},
  \href{http://arxiv.org/abs/1205.0897}{{\normalfont\ttfamily
  arXiv:1205.0897}}\relax
\mciteBstWouldAddEndPuncttrue
\mciteSetBstMidEndSepPunct{\mcitedefaultmidpunct}
{\mcitedefaultendpunct}{\mcitedefaultseppunct}\relax
\EndOfBibitem
\bibitem{LHCb-PAPER-2014-013}
LHCb collaboration, R.~Aaij {\em et~al.},
  \ifthenelse{\boolean{articletitles}}{\emph{{Measurement of $\CP$ asymmetry in
  $\Dz\to\Km\Kp$ and $\Dz\to\pim\pip$ decays}},
  }{}\href{http://dx.doi.org/10.1007/JHEP07(2014)041}{JHEP \textbf{07} (2014)
  041}, \href{http://arxiv.org/abs/1405.2797}{{\normalfont\ttfamily
  arXiv:1405.2797}}\relax
\mciteBstWouldAddEndPuncttrue
\mciteSetBstMidEndSepPunct{\mcitedefaultmidpunct}
{\mcitedefaultendpunct}{\mcitedefaultseppunct}\relax
\EndOfBibitem
\end{mcitethebibliography}

\newpage


\centerline{\large\bf LHCb collaboration}
\begin{flushleft}
\small
R.~Aaij$^{43}$,
B.~Adeva$^{39}$,
M.~Adinolfi$^{48}$,
Z.~Ajaltouni$^{5}$,
S.~Akar$^{59}$,
P.~Albicocco$^{19}$,
J.~Albrecht$^{10}$,
F.~Alessio$^{40}$,
M.~Alexander$^{53}$,
A.~Alfonso~Albero$^{38}$
% ...

{\footnotesize \it
% ...
$ ^{5}$Clermont Universit{\'e}, Universit{\'e} Blaise Pascal, CNRS/IN2P3, LPC, Clermont-Ferrand, France\\
% ...
$ ^{10}$Fakult{\"a}t Physik, Technische Universit{\"a}t Dortmund, Dortmund, Germany\\
% ...
$ ^{19}$Laboratori Nazionali dell'INFN di Frascati, Frascati, Italy\\
% ...
$ ^{38}$ICCUB, Universitat de Barcelona, Barcelona, Spain\\
$ ^{39}$Instituto Galego de F{\'\i}sica de Altas Enerx{\'\i}as (IGFAE), Universidade de Santiago de Compostela, Santiago de Compostela, Spain\\
$ ^{40}$European Organization for Nuclear Research (CERN), Geneva, Switzerland\\
% ...
$ ^{43}$Nikhef National Institute for Subatomic Physics, Amsterdam, The Netherlands\\
% ...
$ ^{48}$H.H. Wills Physics Laboratory, University of Bristol, Bristol, United Kingdom\\
% ...
$ ^{53}$School of Physics and Astronomy, University of Glasgow, Glasgow, United Kingdom\\
% ...
$ ^{59}$University of Cincinnati, Cincinnati, OH, United States\\
% ...
}
\end{flushleft}

\end{document}


{\noindent\bf\Large Supplementary material for LHCb-PAPER-2018-007}

\begin{figure}[b]
\includegraphics[width=8cm]{figs/Fig1a_S.pdf}
\includegraphics[width=8cm]{figs/Fig1b_S.pdf}
\caption{(left) Comparison of the performance of several classifiers, evaluated for the selection
of \decay{\Bm}{\Dsm\Dz} decays, followed by \decay{\Dz}{\Km\pip}.
The two classifiers considered are boosted decision tree (BDT) and multilayer perceptron (MLP).
Several transformations of the input variables are considered: 
I for identity (no transformation),
U for uniform (transform such that the sum of signal and background distribution is uniform),
P for principal component decomposition, and
D for decorrelation.
(right) Output distributions of the BDT for \decay{\Bm}{\Dsm\Dz} decays, followed by \decay{\Dz}{\Km\pip}.
Signal and background distributions are compared and the testing and training samples are overlaid.}
\label{fig:ROC_curve}
\end{figure}

\clearpage

\begin{figure}[t]
\includegraphics[width=12cm]{figs/Fig2_S.pdf}
\caption{Comparison of the BDT response between data and simulation for \decay{\Bm}{\Dsm\Dz} decays, followed by \decay{\Dz}{\Km\pip}.
For the data, the background has been subtracted using sWeights,
where the weight was determined from the reconstructed invariant mass of the \Bm candidate.
}
\label{fig:data_MC_MVA}
\end{figure}

\clearpage

\begin{figure}[tb]
\includegraphics[width=7.5cm]{figs/Fig3a_S.pdf}
\includegraphics[width=7.5cm]{figs/Fig3b_S.pdf}\\
\includegraphics[width=7.5cm]{figs/Fig3c_S.pdf}
\includegraphics[width=7.5cm]{figs/Fig3d_S.pdf}\\
\includegraphics[width=7.5cm]{figs/Fig3e_S.pdf}
\includegraphics[width=7.5cm]{figs/Fig3f_S.pdf}\\
\includegraphics[width=7.5cm]{figs/Fig3g_S.pdf}
\includegraphics[width=7.5cm]{figs/Fig3h_S.pdf}
\caption{Reconstructed invariant-mass distributions of \decay{\Bm}{\DmorDs\Dz} candidates;
left are from simulation and right are from data.
Top row are $\Dsm\Dz$ with \DKpi final states, second row are $\Dsm\Dz$ with \DKpipipi,
third row are $\Dm\Dz$ with \DKpi and on the bottom row are $\Dm\Dz$ with \DKpipipi.
Overlaid are fits to a sum of two Crystal Ball functions, CB1 and CB2, and a background component.
For the simulation, the shape parameters are floating,
while for the data all shape parameters have been fixed to the values obtained from the fit to the simulated events.
}
\label{fig:CB_MC_Bp}
\end{figure}

\clearpage

\begin{table}[b]
  \caption{Yields and raw asymmetries for \decay{\Bm}{\Dsm\Dz} decays, split by \Dz decay mode, data taking year and magnet polarity.}
  \begin{center}\begin{tabular}{lclccc}
  \hline
  Year    & Magnet polarity  & \Dz        & $N(\Bm)$      & $N(\Bp)$      & $A_{\rm raw}$    \\
  \hline
  2011    & \MagUp           & \DKpi      & $\phantom{0} 1649\pm\phantom{0} 45$ & $\phantom{0} 1704\pm\phantom{0} 46$ & $(           -1.6\pm1.9)\%$ \\
          &                  & \DKpipipi  & $\phantom{00} 957\pm\phantom{0} 35$ & $\phantom{00} 978\pm\phantom{0} 37$ & $(           -1.1\pm2.6)\%$ \\
          &                  & combined   & $\phantom{0} 2606\pm\phantom{0} 57$ & $\phantom{0} 2682\pm\phantom{0} 59$ & $(           -1.4\pm1.6)\%$ \\
  \hline                                                                        
  2011    & \MagDown         & \DKpi      & $\phantom{0} 2376\pm\phantom{0} 54$ & $\phantom{0} 2523\pm\phantom{0} 55$ & $(           -3.0\pm1.6)\%$ \\
          &                  & \DKpipipi  & $\phantom{0} 1321\pm\phantom{0} 42$ & $\phantom{0} 1304\pm\phantom{0} 44$ & $(\phantom{-} 0.6\pm2.3)\%$ \\
          &                  & combined   & $\phantom{0} 3696\pm\phantom{0} 68$ & $\phantom{0} 3828\pm\phantom{0} 70$ & $(           -1.7\pm1.3)\%$ \\
  \hline                                                                        
  2011    & combined         & \DKpi      & $\phantom{0} 4024\pm\phantom{0} 71$ & $\phantom{0} 4231\pm\phantom{0} 72$ & $(           -2.5\pm1.2)\%$ \\
          &                  & \DKpipipi  & $\phantom{0} 2286\pm\phantom{0} 55$ & $\phantom{0} 2283\pm\phantom{0} 57$ & $(\phantom{-} 0.1\pm1.7)\%$ \\
          &                  & combined   & $\phantom{0} 6310\pm\phantom{0} 90$ & $\phantom{0} 6513\pm\phantom{0} 92$ & $(           -1.6\pm1.0)\%$ \\
  \hline                                                                        
  2012    & \MagUp           & \DKpi      & $\phantom{0} 4781\pm\phantom{0} 77$ & $\phantom{0} 5073\pm\phantom{0} 79$ & $(           -3.0\pm1.1)\%$ \\
          &                  & \DKpipipi  & $\phantom{0} 2738\pm\phantom{0} 62$ & $\phantom{0} 2958\pm\phantom{0} 64$ & $(           -3.9\pm1.6)\%$ \\
          &                  & combined   & $\phantom{0} 7519\pm\phantom{0} 99$ & $\phantom{0} 8031\pm           101$ & $(           -3.3\pm0.9)\%$ \\
  \hline                                                                        
  2012    & \MagDown         & \DKpi      & $\phantom{0} 4860\pm\phantom{0} 77$ & $\phantom{0} 4870\pm\phantom{0} 78$ & $(           -0.1\pm1.1)\%$ \\
          &                  & \DKpipipi  & $\phantom{0} 2696\pm\phantom{0} 61$ & $\phantom{0} 2711\pm\phantom{0} 59$ & $(           -0.3\pm1.6)\%$ \\
          &                  & combined   & $\phantom{0} 7556\pm\phantom{0} 98$ & $\phantom{0} 7581\pm\phantom{0} 98$ & $(           -0.2\pm0.9)\%$ \\
  \hline                                                                        
  2012    & combined         & \DKpi      & $\phantom{0} 9641\pm           108$ & $\phantom{0} 9979\pm           111$ & $(           -1.7\pm0.8)\%$ \\
          &                  & \DKpipipi  & $\phantom{0} 5430\pm\phantom{0} 87$ & $\phantom{0} 5660\pm\phantom{0} 87$ & $(           -2.1\pm1.1)\%$ \\
          &                  & combined   & $           15071\pm           139$ & $           15639\pm           141$ & $(           -1.8\pm0.6)\%$ \\
  \hline                                                                        
  2011/12 & \MagUp           & \DKpi      & $\phantom{0} 6430\pm\phantom{0} 89$ & $\phantom{0} 6777\pm\phantom{0} 91$ & $(           -2.6\pm1.0)\%$ \\
          &                  & \DKpipipi  & $\phantom{0} 3702\pm\phantom{0} 71$ & $\phantom{0} 3943\pm\phantom{0} 74$ & $(           -3.1\pm1.3)\%$ \\
          &                  & combined   & $           10132\pm           114$ & $           10720\pm           117$ & $(           -2.8\pm0.8)\%$ \\
  \hline                                                                        
  2011/12 & \MagDown         & \DKpi      & $\phantom{0} 7227\pm\phantom{0} 94$ & $\phantom{0} 7432\pm           102$ & $(           -1.4\pm0.9)\%$ \\
          &                  & \DKpipipi  & $\phantom{0} 4019\pm\phantom{0} 74$ & $\phantom{0} 4012\pm\phantom{0} 73$ & $(\phantom{-} 0.1\pm1.3)\%$ \\
          &                  & combined   & $           11246\pm           119$ & $           11444\pm           125$ & $(           -0.9\pm0.8)\%$ \\
  \hline                                                                        
  2011/12 & combined         & \DKpi      & $           13659\pm           129$ & $           14209\pm           132$ & $(           -2.0\pm0.7)\%$ \\
          &                  & \DKpipipi  & $\phantom{0} 7717\pm           103$ & $\phantom{0} 7945\pm           104$ & $(           -1.5\pm0.9)\%$ \\
          &                  & combined   & $           21375\pm           165$ & $           22153\pm           168$ & $(           -1.8\pm0.5)\%$ \\
  \hline
  \end{tabular}\end{center}
\label{table:Arawsplit}
\end{table}